\documentclass[journal]{IEEEtran}
\usepackage{color}
\usepackage{array}
\usepackage{verbatim}
\usepackage{multirow}
\usepackage[cmex10]{amsmath}
\usepackage{amssymb}
\usepackage{graphicx}
\usepackage{makecell}
\usepackage[colorinlistoftodos]{todonotes}
\usepackage[normalem]{ulem}
\makeatletter
\usepackage{empheq}
\usepackage{array}
\usepackage{url}
\usepackage{cite}
\usepackage{units}
\usepackage{siunitx}
\usepackage{epsfig}
\usepackage{xcolor}
\usepackage[caption=false,font=footnotesize]{subfig}
\def\checkmark{\tikz\fill[scale=0.4](0,.35) -- (.25,0) -- (1,.7) -- (.25,.15) -- cycle;}
\providecommand{\tabularnewline}{\\}
\usepackage{bbm}
\usepackage{subfig}
\usepackage{tabularx}
\usepackage{tabu}
\usepackage{enumitem}
\makeatother

\newcommand{\sujan}[1]{\textcolor{blue}{SUJAN: #1}}
\newcommand{\QSArch}[0]{QS-Arch}
\newcommand{\QRArch}[0]{QR-Arch}

\newcommand{\SQNRqiy}{\text{SQNR}_{q_{iy}}}
\newcommand{\SQNRqy}{\text{SQNR}_{q_{y}}}

\newcommand{\SQNRqiydb}{\text{SQNR}_{q_{iy}\text{(dB)}}}
\newcommand{\SQNRqydb}{\text{SQNR}_{q_{y}\text{(dB)}}}

\newcommand{\SNRT}{\text{SNR}_\text{T}}
\newcommand{\SNRs}{\text{SNR}^*_\text{T}}
\newcommand{\SNRa}{\text{SNR}_{\text{a}}}
\newcommand{\SNRA}{\text{SNR}_{\text{A}}}
\newcommand{\SNRTdb}{\text{SNR}_\text{T(dB)}}
\newcommand{\SNRsdb}{\text{SNR}^*_\text{T(dB)}}
\newcommand{\SNRadb}{\text{SNR}_{\text{a(dB)}}}
\newcommand{\SNRAdb}{\text{SNR}_{\text{A(dB)}}}

\newcommand{\yo}{y_{\text{o}}}
\newcommand{\etaa}{\eta_{\text{a}}}
\newcommand{\etae}{\eta_{\text{e}}}
\newcommand{\etah}{\eta_{\text{h}}}

\newcommand{\bympc}{B_y^{\text{MPC}}}

\begin{document}

\bstctlcite{IEEEexample:BSTcontrol}

\title{Fundamental Limits on Energy-Delay-Accuracy of In-memory  Architectures in Inference Applications}

\author{Sujan K. Gonugondla~\IEEEmembership{Member,~IEEE}, Charbel Sakr~\IEEEmembership{Graduate Student Member,~IEEE}, \\
Hassan Dbouk~\IEEEmembership{Student Member,~IEEE}, 
and Naresh R. Shanbhag~\IEEEmembership{Fellow,~IEEE}
\thanks{Charbel Sakr, Hassan Dbouk, and Naresh R. Shanbhag and  are with the Department of Electrical and Computer Engineering, at the University of Illinois at Urbana-Champaign. 

Sujan K. Gonugondla is with Amazon in Seattle, Washington. This work was done during his time at the University of Illinois at Urbana-Champaign.

This work was supported by C-BRIC, one of six centers in JUMP, a
Semiconductor Research Corporation (SRC) program sponsored by DARPA. We thank Professor Boris Murmann for helpful discussions and suggestions.
 }
}

%% article.

\maketitle

\begin{abstract}
This paper obtains fundamental limits on the computational precision of in-memory computing architectures (IMCs). An IMC noise model and associated SNR metrics are defined and their interrelationships analyzed to show that the accuracy of IMCs is fundamentally limited by the compute SNR ($\SNRa$) of its analog core, and that activation, weight and output precision needs to be assigned appropriately for the final output SNR $\SNRT\rightarrow\SNRa$. The minimum precision criterion (MPC) is proposed to minimize the ADC precision. Three in-memory compute models - charge summing (QS), current summing (IS) and charge redistribution (QR) - are shown to underlie most known IMCs. Noise, energy and delay expressions for the compute models are developed and employed to derive expressions for the SNR, ADC precision, energy, and latency of IMCs. The compute SNR expressions are validated via Monte Carlo simulations in a $\unit[65]{nm}$ CMOS process. For a 512 row SRAM array, it is shown that: 1) IMCs have an upper bound on their maximum achievable $\SNRa$ due to constraints on energy, area and voltage swing, and this upper bound reduces with technology scaling for QS-based architectures; 2) MPC enables $\SNRT\rightarrow\SNRa$ to be realized with minimal ADC precision; 3) QS-based (QR-based) architectures are preferred for low (high) compute SNR scenarios.
\end{abstract}

\section{Introduction}

In-memory computing (IMC) \cite{kang2014icassp,CM_patent,kang2020deep,verma2019memory}
has emerged as an attractive alternative to conventional von Neumann (digital) architectures for addressing the energy and latency cost of memory accesses in data-centric machine learning workloads. IMCs embed analog mixed-signal computations in close proximity to the bit-cell array (BCA) in order to execute machine learning computations such as matrix-vector multiply (MVM) and dot products (DPs) as an intrinsic part of the read cycle and thereby avoid the need to access raw data.

IMCs exhibit a fundamental trade-off between its energy-delay product (EDP) and the accuracy or \emph{signal-to-noise ratio} (SNR) of its analog computations. This trade-off arises due to constraints on the maximum bit-line (BL) voltage discharge and due to process variations, specifically spatial variations in the threshold voltage $V_{\text{t}}$, which limit the dynamic range and the SNR. Additionally, IMCs also exhibit noise due to the quantization of its input activation and weight parameters and due to the column analog-to-digital converters (ADCs). Henceforth, we use ``compute SNR" to refer to the computational precision/accuracy of an IMC, and ``precision" to the number of bits assigned to various signals.

Today, a large number of IMC prototype ICs and designs have been demonstrated  \cite{zhang_verma2017jssc,kang2018jssc,jiang2018vlsi,biswas2018conv,gonugondla2018jssc,dbouk2020keyram,khwa201865nm,valavi2018vlsi,kim2019vlsi,dong2020isscc,su2020isscc,xin2020isscc,jaiswal20188T,9050543,8787897,9061142,8998360,9210113,8698312,9146312,8867863,8648391,8006295}. While these IMCs have shown impressive reductions in the EDP over a von Neumann equivalent with minimal loss in inference accuracy, it is not clear that these gains are sustainable for larger problem sizes across data sets and inference tasks. Unlike digital architectures whose compute SNR can be made arbitrarily high by assigning sufficiently high precision to various signals, IMCs need to contend with both quantization noise as well as analog non-idealities. Therefore, IMCs will have intrinsic limits on their compute SNR. Since the compute SNR trades-off with energy and delay, it raises the following question:
\emph{What are the fundamental limits on the achievable computational precision of IMCs?}

Answering this question is made challenging due to the rich design space occupied by IMCs encompassing a huge diversity of available memory devices, bitcell circuit topologies, circuit and architectural design methods. Today's IMCs tend to employ ad-hoc approaches to assign input and ADC precisions or tend to overprovision its analog SNR in order to emulate the determinism of digital computations.

Recently in \cite{mingu-tcas}, we have attempted to answer the above question for the specific IMC in \cite{kang2018jssc}. A comprehensive analytical understanding of the relationship between precision, compute SNR, energy, and delay across all types of IMCs, is presently missing. This paper fills this gap (preliminary results in \cite{ICCAD_precision}) by: 1) defining compute SNR metrics for IMCs, 2) developing a systematic methodology to obtain a minimum precision assignment for activations, weights and outputs of fixed-point DPs realized on IMCs to meet network accuracy requirements, and 3) employing this methodology to obtain the limits on achievable compute SNR of commonly employed IMC topologies, and quantify their energy vs. accuracy trade-offs.

\color{black}
This paper is organized as follows: Section~\ref{sec:prelims} presents preliminaries related to signal and DP quantization. Section~\ref{sec:imcsnrmetrics} proposes the IMC noise model and associated compute SNR metrics. Section~\ref{sec:compute-models} presents analytical expressions for the compute SNR of three different IMCs. Simulation results quantifying various trade-offs including those between energy and accuracy are presented in Section~\ref{sec:simulations}. Section~\ref{sec:guidelines} summarizes the key takeaways for designing IMCs.
\color{black}

\section{Notation and Preliminaries}
\label{sec:prelims}
\subsection{General Notation}
We employ the term signal-to-quantization noise ratio (SQNR) when \emph{only} quantization noise (denoted as $q$) is involved. The term SNR is employed when analog noise sources are included and use $\eta$ to denote such sources. SNR is also employed when both quantization and analog noise sources are present. 

\subsection{The Additive Quantization Noise Model}\label{subsec:adqm}
Under the additive quantization noise model, a floating-point (FL) signal $x$ quantized to $B_x$ bits is represented as $x_q=x+q_x$, where $q_x$ is the quantization noise assumed to be independent of the signal $x$. 

If $x\in [-x_\text{m},x_\text{m}]$ and  $q_x\sim U[-0.5\Delta_x,0.5\Delta_x]$ where $\Delta_x = x_\text{m}2^{-(B_x-1)}$ is the quantization step size and 
$U[a,b]$ denotes the uniform distribution over the interval $[a,b]$, then the signal-to-quantization noise ratio ($\text{SQNR}_x$) is given by:
\begin{align}
\label{eqn:sqnr_db}
    \text{SQNR}_{x(\text{dB})} &= 10\log_{10}(\text{SQNR}_{x})= 6B_x +4.78 -\zeta_{x(\text{dB})}
\end{align}
where $\text{SQNR}_{x} =\frac{\sigma_{x}^2}{\sigma_{q_x}^2}$, $\sigma_{q_x}^2 = \frac{\Delta_x^2}{12}$, and $\zeta_x (\text{dB}) = 10\log_{10}( \frac{x_\text{m}^2}{\sigma_x^2})$ is the peak-to-average (power) ratio (PAR) of $x$. Equation (\ref{eqn:sqnr_db}) quantifies the familiar $\unit[6]{dB}$ SQNR gain per bit of precision.

\subsection{The Dot-Product (DP) Computation}
\label{subsec:dpc}
Consider the FL dot product (DP) computation defined as:
\begin{align}
y_{\text{o}}=\mathbf{w}^{\mathsf{T}}\mathbf{x}=\sum_{j=1}^{N} w_jx_j\label{eqn:DP}
\end{align}
where $y_{\text{o}}$ is the DP of two $N$-dimensional real-valued vectors $\mathbf{w}=[w_1,\ldots,w_N]^{\mathsf{T}}$ (weight vector) and  $\mathbf{x}=[x_1,\ldots,x_N]^{\mathsf{T}}$ (activation vector).

In DNNs, the dot product in (\ref{eqn:DP}) is computed with $w\in [-w_\text{m},w_\text{m}]$ (signed weights), input $x\in [0,x_\text{m}]$ (unsigned activations assuming the use of ReLU activation functions) and output $y\in [-y_\text{m},y_\text{m}]$ (signed outputs). Assuming the additive quantization noise model from Section~\ref{subsec:adqm}, the fixed-point (FX) computation of the DP (\ref{eqn:DP}) with precisions $B_w$ (weight), $B_x$ (activation), and $B_y$ (output), is described by:
\begin{align}
    y &= 
    \mathbf{w}_q^\mathsf{T}\mathbf{x}_q +q_y = (\mathbf{w}+\mathbf{q}_{w})^\mathsf{T} (\mathbf{x}+\mathbf{q}_{x}) + q_y\\
    &\approx\mathbf{w}^{\mathsf{T}}\mathbf{x}+\mathbf{w}^{\mathsf{T}}\mathbf{q}_x+\mathbf{q}_w^{\mathsf{T}}\mathbf{x}+q_y = y_\text{o} + q_{iy}+q_y
    \label{eqn:dp_in_fx}
\end{align}
where $\mathbf{w}_q=\mathbf{w}+\mathbf{q}_{w}$ and $\mathbf{x}_q=\mathbf{x}+\mathbf{q}_{x}$ are the quantized weight and activation vectors, respectively, $q_{iy}$ is the total input (weight and activation) quantization noise seen at the output $y$ (output referred quantization noise), and $q_y$ is the additional output quantization noise due to round-off/truncation in digital architectures or from the finite resolution of the column ADCs in IMC architectures.

Assuming that the weights (signed) and inputs (unsigned) are i.i.d. random variables (RVs), the variances of signals in (\ref{eqn:dp_in_fx}) are given by: 
\begin{align}
      \sigma_{y_{\text{o}}}^2 &= N\sigma^2_w\mathbb{E}[x^2];
      \sigma^2_{q_y} = \frac{\Delta^2_y}{12};
       \sigma^2_{q_{iy}}= \frac{N}{12}\left(\Delta_w^2\mathbb{E}[x^2]  + \Delta_x^2\sigma_w^2\right)
      \label{eqn:quant-noise}
\end{align}
where $\sigma^2_{w}$ is the variance of the weights,  $\Delta_w = w_\text{m}2^{-B_w+1}$, $\Delta_x = x_\text{m}2^{-B_x}$ and $\Delta_y = y_\text{m}2^{-B_y+1}$ are the weight, activation, and output quantization step-sizes, respectively.

\section{Compute SNR Limits of IMCs}
\label{sec:imcsnrmetrics}
\begin{figure}
    \centering
    \includegraphics[width=0.98\linewidth]{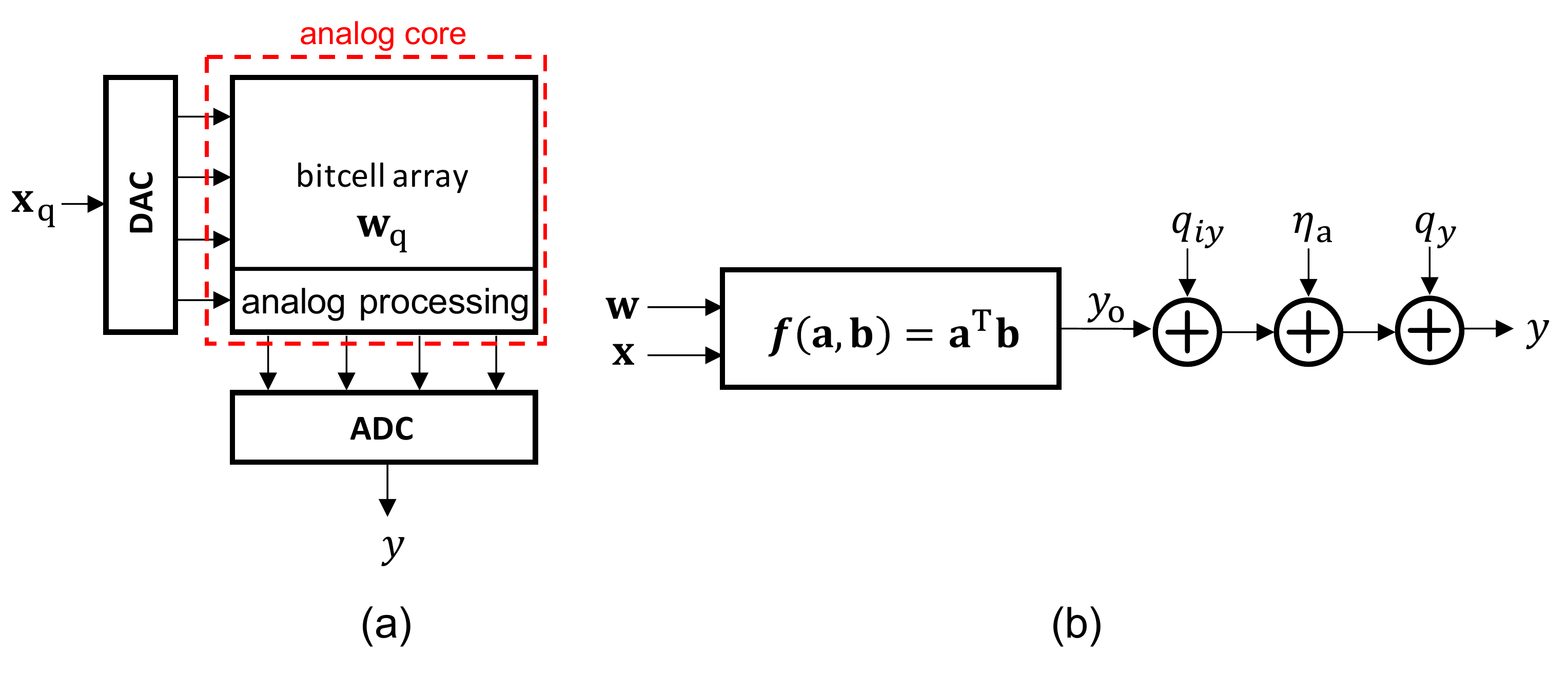}\vspace*{-0.35cm}
    \caption{System noise model of IMC: (a) a generic IMC block diagram, and (b) dominant noise sources in a fixed-point DP computation on IMCs.}
    \label{fig:snr-type}
\end{figure}
%In contrast to digital architectures, IMCs (Fig.~\ref{fig:snr-type}) exhibit four dominant noise sources:
We propose the system noise model in Fig.~\ref{fig:snr-type} for obtaining precision limits on IMCs. Such architectures (Fig.~\ref{fig:snr-type}(a)) accept a quantized input ($\mathbf{x}_q$) and a quantized weight vector ($\mathbf{w}_q$) to implement multiple FX DP computations of \eqref{eqn:dp_in_fx} in parallel in its analog core. Hence, unlike digital architectures, IMC architectures suffer from both quantization and analog noise sources such as SRAM cell current variations, thermal noise, and charge injection, as well as the limited headroom, which limits its compute SNR. 

%Therefore, the use of MPC becomes critical in ensuring that IMC architectures are able to maximize their accuracy under their analog constraints. 

%\begin{itemize}[leftmargin=*]
%    \item data (weight and activation) quantization noise $\eta_{\text{q}}$ (Fig.~\ref{fig:snr-type}(b)).
%    \item distortion $\eta_{\text{h}}$ from clipping of $y_{\text{a}}$ caused by the limited voltage swing permitted on the BLs shown as a saturating non-linearity $g()$ (Fig.~\ref{fig:snr-type}(b)).
%    \item circuit noise $\eta_{\text{e}}$ which captures the impact of circuit non-idealities such as spatial variations in transistor threshold voltage $V_t$ (Fig.~\ref{fig:snr-type}(b)).
%\item quantization noise $\eta_{\text{d}}$ due to the finite resolution and the finite input swing of the analog-to-digital converter (ADC)  (Fig.~\ref{fig:snr-type}(c)). It depends upon the input swing and the resolution $B_y$ of the ADC.
%\end{itemize}

\subsection{Compute SNR Metrics for IMCs}
The following equations describe the IMC noise model in Fig.~\ref{fig:snr-type}:
\begin{align}
y & = \yo+q_{iy}+\etaa+q_y;\quad\etaa=\etae+\etah\label{eqn:imc-noise-model}
\end{align}
where $\yo$ is the ideal DP value defined in (\ref{eqn:DP}), $q_{iy}$ is the output referred quantization noise, $\eta_{\text{a}}$ is the analog noise term comprising both clipping noise $\etah$ due to limited headroom and other noise sources $\etae$, and $q_y$ is the output quantization noise introduced by the ADC. 

We define the following fundamental compute SNR metrics:
\begin{align}
    \SQNRqiy &= \frac{\sigma_{\yo}^2}{\sigma_{q_{iy}}^2}; \SNRa = \frac{\sigma_{\yo}^2}{\sigma_{\etaa}^2};\SQNRqy = \frac{\sigma_{\yo}^2}{\sigma_{q_{y}}^2}\label{eqn:imc-accuracy-metrics}
\end{align}
where $\SNRa$ is the \emph{analog SNR}, $\SQNRqiy$ is the \emph{output referred SQNR} due to input (weight and activation) quantization and is given by:
\begin{align}
   \text{SQNR}_{q_{iy}(\text{dB})} &= 6(B_x+B_w)+4.8-[\zeta_{x (\text{dB})}+\zeta_{w(\text{dB})}]\nonumber\\ 
    &\quad 
    -10\log_{10}\left(\frac{2^{2B_x}}{\zeta_x^2}+\frac{2^{2B_w}}{\zeta_w^2} \right) \label{eqn:sqnrqiy}
\end{align}
where $\zeta_{x (\text{dB})} = 10\log_{10}\left(\frac{x_\text{m}^2}{4\mathbb{E}[x^2]}\right)$ and $\zeta_{w (\text{dB})} = 10\log_{10}\left(\frac{w_\text{m}^2}{\sigma_w^2}\right)$ are the PARs of the (unsigned) activations and (signed) weights, respectively, and $\SQNRqy$ is the \emph{digitization SQNR} solely due to ADC quantization noise and is given by:
\begin{align}
 \SQNRqydb &=    6B_y+4.8-[\zeta_{x\text{(dB)}}+\zeta_{w\text{(dB)}}]-10\log_{10}(N)
 \label{eqn:sqnrqy}
\end{align}
obtained by the substitutions: $B_x\leftarrow B_y$ and $\zeta_{x\text{(dB)}}\leftarrow\zeta_{y\text{(dB)}}=\zeta_{x\text{(dB)}}+\zeta_{w\text{(dB)}}+10\log_{10}(N)$ in \eqref{eqn:sqnr_db}.

From \eqref{eqn:imc-noise-model} and \eqref{eqn:imc-accuracy-metrics}, it is straightforward to show:
\begin{align}
    \SNRA & = \frac{\sigma_{\yo}^2}{\sigma_{q_{iy}}^2+\sigma_{\etaa}^2}=\left [\frac{1}{\SNRa}+\frac{1}{\SQNRqiy}\right ]^{-1}\label{eqn:SNRA}\\
    \SNRT & = \frac{\sigma_{\yo}^2}{\sigma_{q_{iy}}^2+\sigma_{\etaa}^2+\sigma_{q_{y}}^2}= \left [\frac{1}{\SNRA}+\frac{1}{\SQNRqy}\right ]^{-1}
    \label{eqn:SNRT}
\end{align}
where $\SNRA$ is the pre-ADC SNR and $\SNRT$ is the total output SNR including all noise sources. Note: \eqref{eqn:SNRA}-\eqref{eqn:SNRT} can be repurposed for digital architectures by setting $\SNRa\rightarrow\infty$ since quantization is the only noise source thereby implying $\SNRA=\SQNRqiy$. 
Equations \eqref{eqn:sqnrqiy}-\eqref{eqn:sqnrqy} indicate that $\SQNRqiy$ and $\SQNRqy$ can be made arbitrarily large by assigning sufficiently high precision to the DP inputs ($B_x$ and $B_w$) and the output ($B_y$). Thus, from \eqref{eqn:SNRA}-\eqref{eqn:SNRT}, $\SNRT$ in IMCs is fundamentally limited by $\SNRa$ which depends on the analog noise sources as one expects. 
\subsection{Maximizing $\SNRT$ in IMCs}
\label{subsec:prec_methodology}
\begin{figure}[!t]
    \centering
    \includegraphics[width=0.85\linewidth,trim=0.1cm 0.1cm 0.1cm 0.1cm,clip]{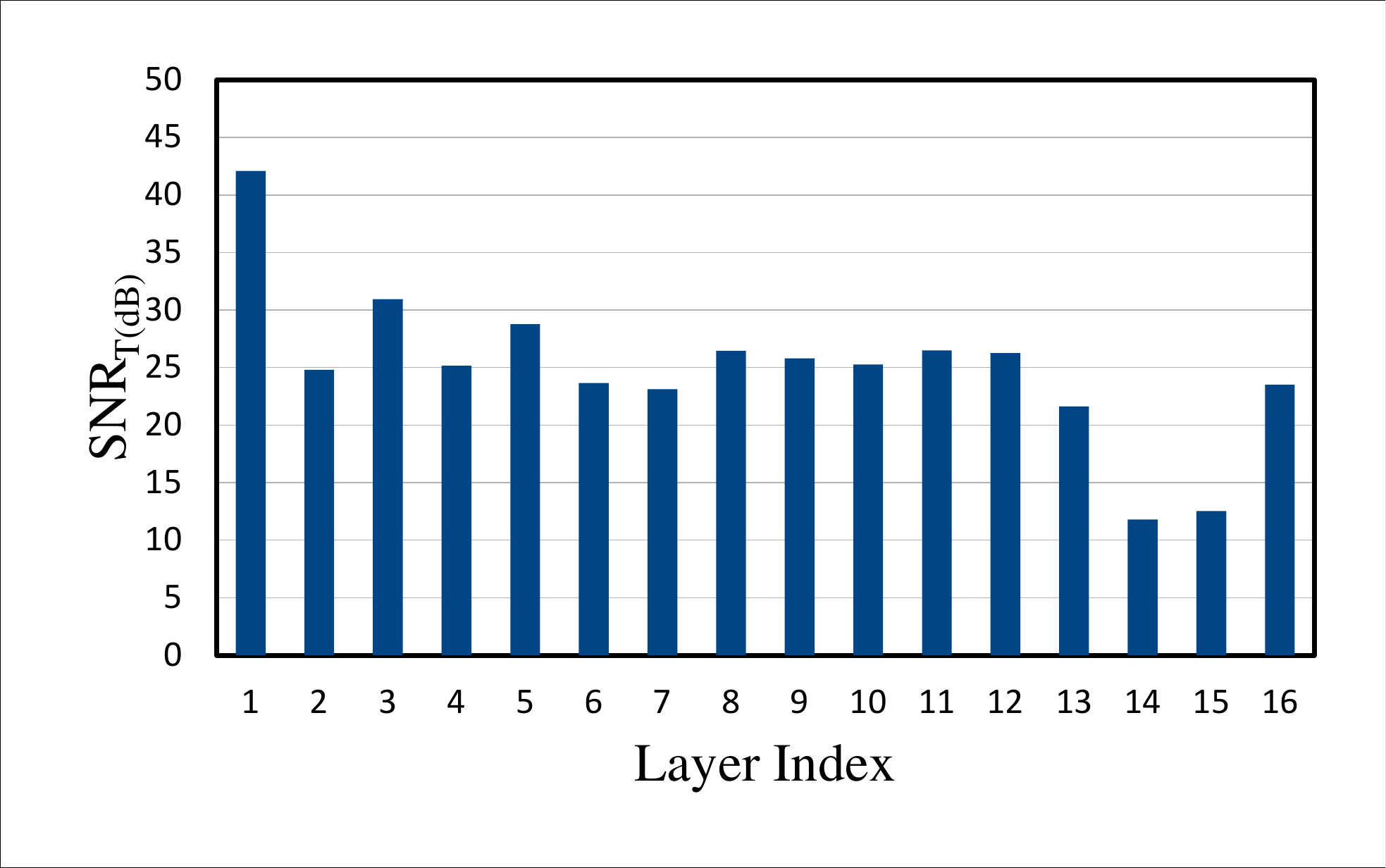}
    \caption{Per-layer $\SNRTdb$ requirements of DP computations in VGG-16 deployed on ImageNet.}
    \label{fig:vgg-snr}
\end{figure}
Prior work based on post-training quantization \cite{sakr2017icml,sakr2018icassp}, indicates the requirement $\SNRTdb>\SNRsdb=\unit[10]{dB}$-$\unit[40]{dB}$ (see Fig.~\ref{fig:vgg-snr}) for the inference accuracy of an FX network to be within 1\% of the corresponding FL network for popular DNNs (AlexNet, VGG-9, VGG-16, ResNet-18) deployed on the ImageNet and CIFAR-10 datasets. While in-training quantization methods \cite{hubara2016binarized} can reduce these requirements, a precision of 4-b ($\sim\unit[24]{dB}$) is generally found to be \cite{choi2018pact} sufficient. To meet this $\SNRTdb$ requirement, digital architectures choose $B_x$ and $B_w$ such that $\SQNRqiy>\SNRs$, and then choose $B_y$ sufficiently high to guarantee $\SQNRqy\gg\SQNRqiy$ so that $\SNRT\rightarrow\SQNRqiy$.

In contrast, for IMCs, we first need to ensure that $\SNRa>\SNRs$ so that $\SNRT$ can be made to approach $\SNRa$ with appropriate precision assignment. Such a precision assignment can be easily derived from \eqref{eqn:SNRA}-\eqref{eqn:SNRT} as shown below:
\begin{enumerate}
    \item Assign sufficiently high values for $B_x$ and $B_w$ per \eqref{eqn:sqnrqiy} such that $\SQNRqiy\gg\SNRa$ so that $\SNRA\rightarrow\SNRa$ per \eqref{eqn:SNRA}.
    \item Assign sufficiently a high value for $B_y$ such that $\SQNRqy\gg\SNRA$ so that $\SNRT\rightarrow\SNRA$ per \eqref{eqn:SNRT}.
\end{enumerate}
For example, if $\SQNRqiydb,\SQNRqydb\geq\SNRadb + \unit[9]{dB}$ then  $\SNRadb-\SNRTdb\leq \unit[0.5]{dB}$, i.e., $\SNRTdb$ lies within $\unit[0.5]{dB}$ of $\SNRadb$. In this manner, by appropriate choices for $B_x$, $B_w$, and $B_y$, IMCs can be designed such that $\SNRT\rightarrow\SNRa$, which, as mentioned earlier, is the fundamental limit on $\SNRT$.

From the above discussion it is clear that the input precisions $B_x$ and $B_w$ are dictated by network accuracy requirements, while the output precision $B_y$ needs to be set sufficiently high for the output quantization from becoming a significant noise contributor. To ensure that a sufficiently high value for $B_y$, digital architectures employ the \emph{bit growth criterion} (BGC) described next.

\subsection{Bit Growth Criterion (BGC)}
\label{subsec:bgc}
The BGC is commonly employed to assign the output precision $B_y$ in digital architectures \cite{guptaICML,sakr2017icml}. BGC sets $B_y$ as:
\begin{align}
    B_y^{\text{BGC}} &= B_x + B_w + \log_2(N) \label{eqn:bgc_assignment}
\end{align}
Substituting $B_y=B_y^{\text{BGC}}$ from \eqref{eqn:bgc_assignment} into \eqref{eqn:sqnrqy} and employing the relationship $\zeta_{y\text{(dB)}}=10\log_{10}(N)+\zeta_{x\text{(dB)}}+\zeta_{w\text{(dB)}}$, the resulting SQNR due to output quantization using the BGC is given by:
\begin{align}
    ~&\SQNRqydb^{\text{BGC}} =10\log_{10} \left(\frac{\sigma_{\yo}^2}{\sigma^2_{q_y}}\right) \nonumber\\
    &= 6(B_x+B_w)+4.8-[\zeta_{x (\text{dB})}+\zeta_{w(\text{dB})}]+10\log_{10}(N).\label{eqn:sqnr-bgc}
\end{align}
Recall that $\SQNRqy^{\text{BGC}}\gg \SNRA$ in order to ensure $\SNRT$ is close to its upper bound. Comparing \eqref{eqn:sqnrqy} and \eqref{eqn:sqnr-bgc}, we see that, for high values of DP dimensionality $N$, BGC is overly conservative since it assigns large values to $B_y$ per \eqref{eqn:bgc_assignment}. 
%In fact, \eqref{eqn:sqnrqi2y} and \eqref{eqn:sqnr-bgc} differ only in their last ($10\log_{10}(\ldots)$) term, whereby in \eqref{eqn:sqnrqi2y}, this term is independent of $N$ and has a negative sign whereas the reverse is true for \eqref{eqn:sqnr-bgc}. 
Some digital architectures truncate the LSBs to control bit growth. The SQNR of such \emph{truncated BGC} (tBGC) can be obtained directly from \eqref{eqn:sqnrqy} by setting the value of $B_y<B_y^{\text{BGC}}$.

BGC's high precision requirements is accommodated in digital architectures by increasing the precision of arithmetic units with a commensurate increase in the computational energy, latency, and activation storage costs. However, IMCs cannot afford to use this criterion since $B_y$ is the precision of the BL ADCs which impacts its energy, latency, and area. Indeed, recent works \cite{dally2020dac} have claimed that BL ADCs dominate the energy and latency costs of IMCs. Such works employ the highly conservative BGC or tBGC to assign $B_y$. 

In the next section, we propose an alternative to BGC and tBGC referred to the \emph{minimum precision criterion} (MPC), that can be employed by both digital and IMC architectures to achieve a desired $\SQNRqy$ with much smaller values of $B_y$.

\subsection{The Minimum Precision Criterion (MPC)}
\label{subsec:mpc}
\begin{figure}[!t]
    \centering
    \subfloat[]{
    \includegraphics[width=0.49\linewidth,trim=0.1cm 0.1cm 0.1cm 0.1cm,clip,page=3]{Figures/distributions_comparison.pdf}
    }
    \subfloat[]{
    \includegraphics[width=0.49\linewidth,trim=0.1cm 0.1cm 0.1cm 0.1cm,clip,page=2]{Figures/distributions_comparison.pdf}
    }\\
    \subfloat[]{
    \includegraphics[width=0.49\linewidth,trim=0.1cm 0.1cm 0.1cm 0.1cm,clip,page=1]{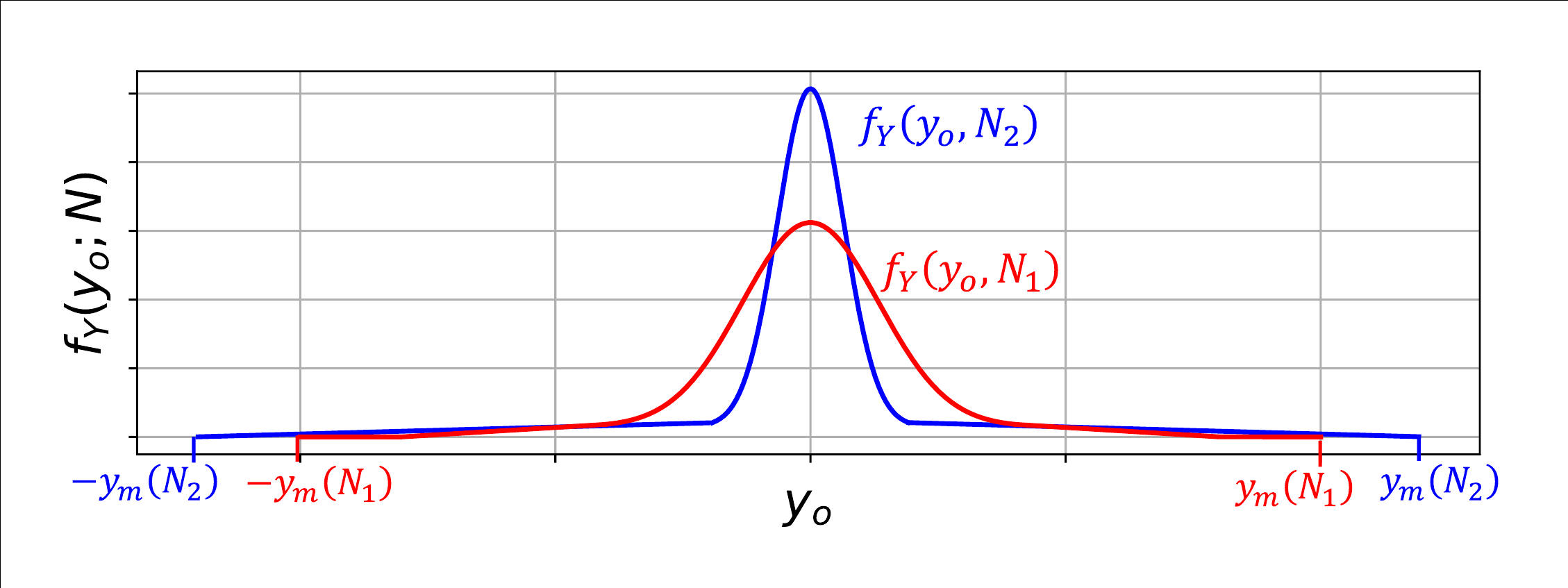}
    }
    \caption{Comparison of BGC and MPC: (a) MPC quantization levels, (b) BGC quantization levels, and (c) distribution $f_Y(y_\text{o})$ of the ideal DP output $y_\text{o}$ vs. DP dimensionality $N$.}
    \label{fig:bgc-mpc-lm}
\end{figure}

We propose MPC to reduce $B_y$ without incurring any loss in $\SQNRqy$ compared to BGC. \color{black} Unlike BGC, MPC accounts for the statistics of $y_\text{o}$ 
to permit controlled amounts of \emph{clipping} to occur. In MPC (see Fig.~\ref{fig:bgc-mpc-lm}(a)), the output $y_\text{o}$ is \emph{clipped} to lie in the range $[-y_\text{c},y_\text{c}]$ instead of $[-y_\text{m},y_\text{m}]$ as in BGC (see Fig.~\ref{fig:bgc-mpc-lm}(b)), where $y_\text{c}<y_\text{m}$ is the \emph{clipping level}, and $B_y$ bits are employed to quantize this reduced range. The \emph{clipping probability} $p_{\text{c}}=\Pr\{|y_o|>y_\text{c}\}$ is kept to a small user-defined value, e.g., $y_\text{c}=4\sigma_{y_\text{o}}$ ensures that $p_\text{c}<0.001$ if $y_\text{o}\sim\mathcal{N}(0,\sigma_{y_\text{o}}^2)$. The resulting $\text{SQNR}_{q_y}$ is given by:
\begin{align}
   \SQNRqydb^{\text{MPC}}  =& 6B_y + 4.8 - \zeta_{y\text{(dB)}}^{\text{MPC}}-10\log_{10}\left(1+p_\text{c}\frac{\sigma^2_{cc}}{\sigma^2_{q_y}}\right)\label{eqn:sqnr-mpc}
\end{align}
%\begin{align}
%    \text{SQNR}_{y-\text{MPC}} & = 6B_y + 4.78 - 20\log(\zeta)\nonumber \\&- 10\log\left(1+p\frac{3y_c^2\mathbbm{E}\left[\left(y_{\text{o}}-y_c\right)^2\bigg||y_{\text{o}}|>y_c\right]}{2^{(2B_y-4)}}\right)\label{eqn:sqnr_mpc_tradeoff}
%\end{align}
where $\zeta_{y(\text{dB})}^{\text{MPC}} = 10\log_{10}\left(\frac{y_\text{c}^2}{\sigma_{\yo}^2}\right)$, and $\sigma^2_{cc} =\mathbb{E}\big[\left(y_{\text{o}}-y_\text{c}\right)^2\big||y_{\text{o}}|>y_\text{c}\big]$ is the \emph{conditional clipping noise variance}. Setting $y_\text{c} = \zeta_y^{\text{MPC}} \sigma_{y_o}$ yields $\zeta_{y\text{(dB)}}^{\text{MPC}} = 10\log_{10}(\zeta_y^{\text{MPC}})^2$ indicating that $p_\text{c}$ is a decreasing function of $\zeta_y^{\text{MPC}}$.
%For example, if $y_\text{o}\sim\mathcal{N}(0,\sigma_{y_\text{o}}^2)$, we have $p_\text{c}=Q(\zeta_y^{\text{MPC}})$, where $Q(\ldots)$ is the complementary CDF of a standard normal distribution. 
Thus, \eqref{eqn:sqnr-mpc} has the same form as \eqref{eqn:sqnr_db} with an additional (last term) \emph{clipping noise factor}.

MPC exploits a key insight (see Fig.~\ref{fig:bgc-mpc-lm}(c)), which follows from the Central Limit Theorem (CLT) -- \emph{in a $N$-dimensional DP computation (\ref{eqn:DP}), $\sigma_{y_\text{o}}$ grows sub-linearly (as $\sqrt{N}$) as compared to the maximum $y_\text{m}$ which grows linearly with $N$}. 
Furthermore, \eqref{eqn:sqnr-mpc} shows a \emph{quantization vs. clipping noise trade-off} controlled by the clipping level $y_\text{c}$. We show empirically in Section~\ref{subsec:mpc-sim} that $\SQNRqydb^{\text{MPC}}$ in \eqref{eqn:sqnr-mpc} is maximized when clipping level $y_\text{c}=4\sigma_{y_\text{o}}$ if $y_\text{o}\sim\mathcal{N}(0,\sigma_{y_\text{o}}^2)$. This trade-off (see Fig.~\ref{fig:bgc-mpc-lm}(b)) is absent in BGC and tBGC, and is critical to MPC's ability to realize desired values of $\SQNRqy$ with smaller values of $B_y$.

Thus, we state the following MPC-based rule for maximizing the SQNR of column ADCs in IMCs:
\begin{quote}
\begin{center} \textbf{MPC-based SQNR Maximizing Rule}\end{center}
    \emph{For a Gaussian signal, setting the clipping level to four times the standard deviation will maximize the SQNR for a given precision $B_y$.}
\end{quote}
Some IMC designs \cite{seo_mpc} do in fact allow for clipping in the column ADCs but these levels are set empirically. The MPC-based Rule in contrast quantitatively specifies the smallest ADC precision and the optimal clipping level needed to ensure that $\SQNRqy\gg\SNRA$.
%Architects can also employ this rule to minimize the precision of the partial product accumulator in digital DNN accelerators.
A lower bound on $B_y$ can be obtained by assuming $y_\text{o}\sim\mathcal{N}(0,\sigma_{y_\text{o}}^2)$, and substituting $y_\text{c}=4\sigma_{y_\text{o}}$, and $p_\text{c}=0.001$ into \eqref{eqn:sqnr-mpc}, to obtain:
\begin{align}
   \bympc \geq \frac{1}{6}\left[\SNRAdb + 7.2 - \gamma -10\log_{10}\left(1-10^{-\frac{\gamma}{10}} \right) \right] %\frac{\text{SQNR}_{y\text{(dB)}}^{\text{MPC}}+16.2}{6}
    \label{eqn:bympc}
\end{align}
in order for $\SNRAdb-\SNRTdb\leq \gamma$. For instance, the choice $\gamma = \unit[0.5]{dB}$ yields $\bympc \geq \frac{1}{6}\left[\SNRAdb + 16.3 \right]$ which corresponds to $\text{SQNR}_{y\text{(dB)}}^{\text{MPC}} \geq \SNRAdb + \unit[9]{dB}$ as discussed in Section \ref{subsec:prec_methodology}.
%This implies that $\unit[19]{dB}\leq \text{SNR}_{\text{d}}\leq \unit[49]{dB}$ for minimal impact on inference accuracy compared to a FL network. This implies that the column ADC precision in IMCs needs to be between 4-b to 8-b. Unsurprisingly, most IMCs (see Table~\ref{tab:classify}) today do implement 1-b to 8-b ADCs though many times these precision values are chosen based on implementation constraints.
\subsection{Simulation Results}
\label{subsec:mpc-sim}
\begin{figure}[!t]
    \centering
    %\hspace*{-0.4cm}
     \subfloat[]{
    \includegraphics[width=0.6\linewidth,trim=0.15cm 0.15cm 0.15cm 0.15cm,clip,page=2]{Figures/mpc_sqnrs.pdf}
    }\\
    %\hspace*{-0.4cm}
    \subfloat[]{
    \includegraphics[width=0.6\linewidth,trim=0.15cm 0.15cm 0.15cm 0.15cm,clip,page = 1]{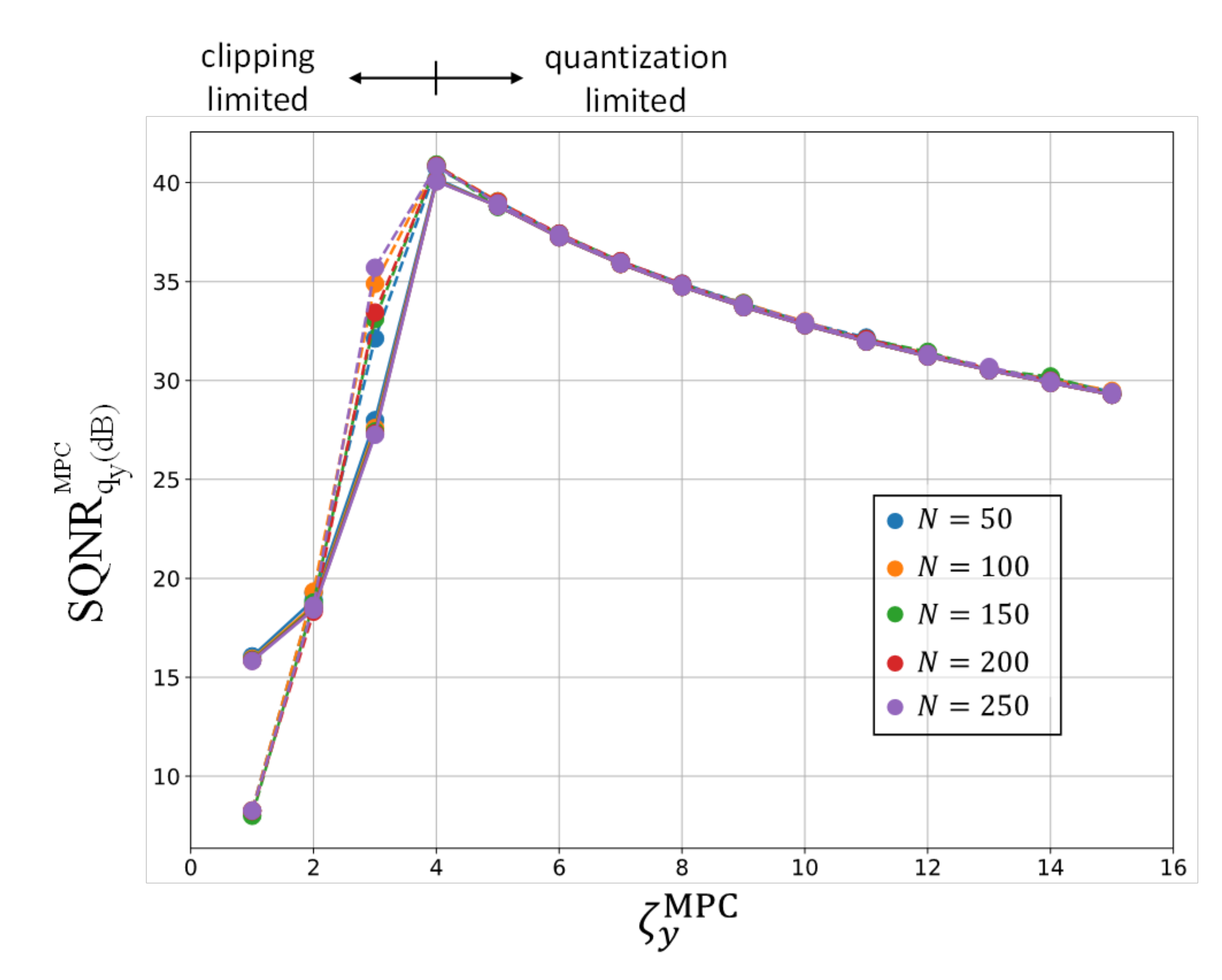}
    }
    \caption{Trends in $\SQNRqydb$ for DP computation with $B_x=B_w=7$: (a) $\SQNRqydb$ vs. $N$ for MPC ($\zeta_y=4)$, BGC, tBGC,
    and (b) $\SQNRqydb^{\text{MPC}}$ vs. $\zeta_y^{\text{MPC}}$ when $B_y=8$.
    %Numerical evaluation of expressions (\eqref{eqn:sqnrqy}, \eqref{eqn:sqnr-bgc},  \eqref{eqn:sqnr-mpc}) (bold) and Monte Carlo simulations (dotted) match well. 
    }
    \label{fig:mpc-sqnr}
\end{figure}
To illustrate the difference between MPC, BGC and tBGC, we assume that $\SNRadb\geq\unit[31]{dB}$, so that $\SNRTdb\geq \unit[30]{dB}$ provided $\SQNRqiydb,\SQNRqydb\geq\unit[40]{dB}$ per \eqref{eqn:SNRA}-\eqref{eqn:SNRT}. We further assume DPs of varying dimension $N$ with 7-b quantized unsigned inputs and signed weights randomly sampled from uniform distributions. Substituting $B_x=B_w=7$,  $\zeta_{x\text{(dB)}}=\unit[-1.3]{dB}$, and $\zeta_{w\text{(dB)}}=\unit[4.8]{dB}$ into \eqref{eqn:sqnrqiy}, we obtain $\SQNRqiydb=\unit[41]{dB}$. Thus, all that remains is to assign $B_y$ such that $\SQNRqydb\geq\unit[40]{dB}$, for which there are three choices - MPC, BGC and tBGC. 

Figure~\ref{fig:mpc-sqnr}(a) compares the $\SQNRqy$ achieved by the three methods. Per \eqref{eqn:bympc}, MPC meets the $\SQNRqydb\geq\unit[40]{dB}$ requirement by setting $B_y=8$ and $\zeta_{y}^{\text{MPC}}=4$ independent of $N$. In contrast, per \eqref{eqn:bgc_assignment}, BGC assigns $16\leq B_y\leq 20$ as a monotonically increasing function of $N$ to achieve the same $\SNRT$ as MPC. Furthermore, tBGC meets the $\SQNRqy$ requirement with  $11\leq B_y\leq 13$ but fails to do so with $B_y=8$.
Figure~\ref{fig:mpc-sqnr}(b) shows that $\SQNRqydb^{\text{MPC}}$ is maximized when $\zeta_{y}^{\text{MPC}}=4$, i.e., when clipping level $y_\text{c}=4\sigma_{y_\text{o}}$ thereby illustrating 
MPC's quantization vs. clipping noise trade-off described by \eqref{eqn:sqnr-mpc}. 

Figure~\ref{fig:mpc-sqnr} also validates the analytical expressions \eqref{eqn:sqnrqiy}, \eqref{eqn:sqnrqy}, \eqref{eqn:sqnr-bgc}, and \eqref{eqn:sqnr-mpc} (bold) by indicating a close match to ensemble-averaged values of $\SQNRqy$ obtained from Monte Carlo simulations (dotted). 

Note: it is well-established that the theoretically optimal quantizer given an arbitrary signal distribution is obtained from the Lloyd-Max (LM) algorithm \cite{lloyd_max}.  Unfortunately, the LM quantization levels are non-uniformly spaced which makes it hard to design efficient arithmetic units to process such signals. Furthermore, for $B_y=8$ as in Figure~\ref{fig:mpc-sqnr}(b), LM achieves an $\SQNRqydb=\unit[41.31]{dB}$ which is only $\unit[0.5]{dB}$ better than MPC. Thus, MPC offers a practical alternative to LM for assigning minimal precision to the column ADC in IMCs and the accumulator precision in digital architectures.
\section{Analytical Models for Compute SNR}
\label{sec:compute-models}
\begin{table}
    %\centering
    \caption{A Taxonomy of CMOS IMC Designs using In-memory Compute Models \label{tab:classify}
    }
	\resizebox{0.97\linewidth}{!}{
		\centering
    \begin{tabular}{|c||c|c|c||c|c|c|}
    \hline
         & \multicolumn{3}{c||}{\multirow{2}{*}{\shortstack{In-memory \\ Compute Model}}} & \multicolumn{2}{c|}{\multirow{2}{*}{\shortstack{Analog Core\\ Precision}}} & \multicolumn{1}{c|}{\multirow{2}{*}{\shortstack{ADC \\ Precision}}}  \tabularnewline
         & \multicolumn{3}{c||}{} & \multicolumn{2}{c|}{} & \tabularnewline
         \cline{2-7}
          &  QS &  IS & QR & $B_x$  & $B_w$ & $B_\text{ADC}$ \tabularnewline
         \hline
         \hline 
         Kang \emph{et al.} \cite{kang2018jssc} & \checkmark &  & \checkmark & 8 & 8 & 8\tabularnewline
         
         \hline
          Biswas \emph{et al.} \cite{biswas2018conv} & & & \checkmark & 8 & 1& 7 \tabularnewline
         
         \hline 
          Zhang \emph{et al.} \cite{zhang_verma2017jssc} & \checkmark & & & 5 & 1 & 1\tabularnewline
         
         \hline 
          Valavi \emph{et al.} \cite{valavi2018vlsi}& & & \checkmark & 1 & 1& 1 \tabularnewline
         
         \hline 
         Khwa \emph{et al.} \cite{khwa201865nm} & &\checkmark  & & 1 & 1 & 1\tabularnewline
         
         \hline 
         Jiang \emph{et al.} \cite{jiang2018vlsi} & & \checkmark  & & 1 & 1& 3.46 \tabularnewline
         
         \hline 
         Si \emph{et al.} \cite{si2019isscc} & \checkmark &   & \checkmark & 2 & 5 & 5\tabularnewline
        
         \hline 
         Jia \emph{et al.} \cite{jia2018arxiv} &  & & \checkmark & 1 & 1& 8 \tabularnewline
        
         \hline 
         Okumura \emph{et al.} \cite{okumura2019vlsi} &    & \checkmark& & 1 & T  & 8\tabularnewline
        
         \hline 
         Kim \emph{et al.} \cite{kim2019vlsi} & &\checkmark   &  &1 & 1& 1\tabularnewline
        
         \hline 
         Guo \emph{et al.} \cite{guo2019vlsi} &  \checkmark & &  & 1 & 1 & 3\tabularnewline
         
         \hline 
         Yue \emph{et al.} \cite{yue2020isscc}  & \checkmark &   & \checkmark & 2 & 5 & 5\tabularnewline
         
         \hline 
         Su \emph{et al.} \cite{su2020isscc}  & \checkmark &   &  & 2 & 1 & 5\tabularnewline
         
         \hline 
         Dong \emph{et al.} \cite{dong2020isscc}  & \checkmark &   &  \checkmark & 4 & 4 & 4\tabularnewline
         
         \hline 
          Si \emph{et al.} \cite{xin2020isscc}  & \checkmark &   &   & 2 & 2 & 5\tabularnewline
         
         \hline 
          Jiang \emph{et al.} \cite{jiang2020c3sram}  &  &   & \checkmark  & 1 & 1 & 5\tabularnewline
         
         \hline 
          Jaiswal \emph{et al.} \cite{jaiswal20188T}  &  & \checkmark   &  & 4 & 4 & 4\tabularnewline
         
         \hline 
          Ali \emph{et al.} \cite{9050543}  &\checkmark  &    & \checkmark & 4 & 4 & 4 \tabularnewline
         
         \hline 
          Si \emph{et al.} \cite{8787897}  & \checkmark &  &   & 1 & 1 & 1 \tabularnewline
         
         \hline 
          Liu \emph{et al.} \cite{9061142}  &  & \checkmark &   & A & 1 & 1 \tabularnewline
         
         \hline 
          Zhang \emph{et al.} \cite{8998360}  &  & \checkmark &   & 8 & 8 & 8 \tabularnewline
         
         \hline 
          Gong \emph{et al.} \cite{9210113}  & \checkmark &  &   & 2 & 3 & 8 \tabularnewline
         
         \hline 
          Agrawal \emph{et al.} \cite{8698312}  &  &  & \checkmark  & 1 & 1 & 5 \tabularnewline
         \hline
    \end{tabular}
    }
	\vspace{4pt}
	
	{\footnotesize  T: Ternary; A: Analog/Continuous-valued}
	
\end{table}
This section derives analytical expressions for $\SNRa$ of a typical IMC. We introduce compute models that form the fundamental building blocks of IMCs and present analytical expressions for circuit domain equivalents of $\eta_{\text{e}}$ and $\eta_\text{h}$ in \eqref{eqn:imc-noise-model} for them. These are combined with algorithm and precision-dependent noise sources $q_{iy}$ and $q_y$ to obtain $\SNRT$. First, we show that most IMCs can be `explained' via three in-memory compute models.
\subsection{In-memory Compute Models}
All IMCs are viewed as employing one or more \emph{in-memory compute models} defined as a mapping of algorithmic variables $y_{\text{o}}$, $x_j$ and $w_j$ in (\ref{eqn:DP}) to physical quantities such as time, charge, current, or voltage, in order to (usually partially) realize an analog BL computation of the multi-bit DP in (\ref{eqn:DP}).  

Furthermore, we suggest that most IMCs today employ one or more of the following three in-memory compute models (see Fig.~\ref{fig:compute-models}): (a) \emph{charge summing} (QS) \cite{kang2018jssc,kang2015icassp,gonugondla2018jssc,zhang_verma2017jssc}; (b) \emph{current summing} (IS) \cite{kim2019vlsi,jiang2018vlsi,khwa201865nm,si2019isscc}; and (c) \emph{charge redistribution} (QR) \cite{valavi2018vlsi,biswas2018conv,kang2018jssc, gonugondla2018jssc}, and conjecture that these compute models are in some sense universal in that they represent an approximation to a `complete set' of practical, i.e., realizable, mappings of variables from the algorithmic to the circuit domain as shown in Table~\ref{tab:classify}. 

Henceforth, we discuss the QS model and QR compute model and the corresponding IMC architectures referred to as \QSArch{}, \QRArch{} in detail since it is very commonly used. We also study compute-memory (CM) architectures \cite{kang2014icassp,kang2018jssc,gonugondla2018jssc} which combine the QS and QR compute models to implement a multi-bit DP.

Table~\ref{tab:param} tabulates  parameters of the QS and QR models in a representative \unit[65]{nm} CMOS process. Table~\ref{tab:arch_summary} summarizes the attributes of all three architectures.
\subsection{Charge Summing (QS)}\label{subsec:QS-model}

\subsubsection{The QS Model}
\begin{figure}[!t]
    \centering
    \subfloat[]{
    \includegraphics[width=0.36\linewidth]{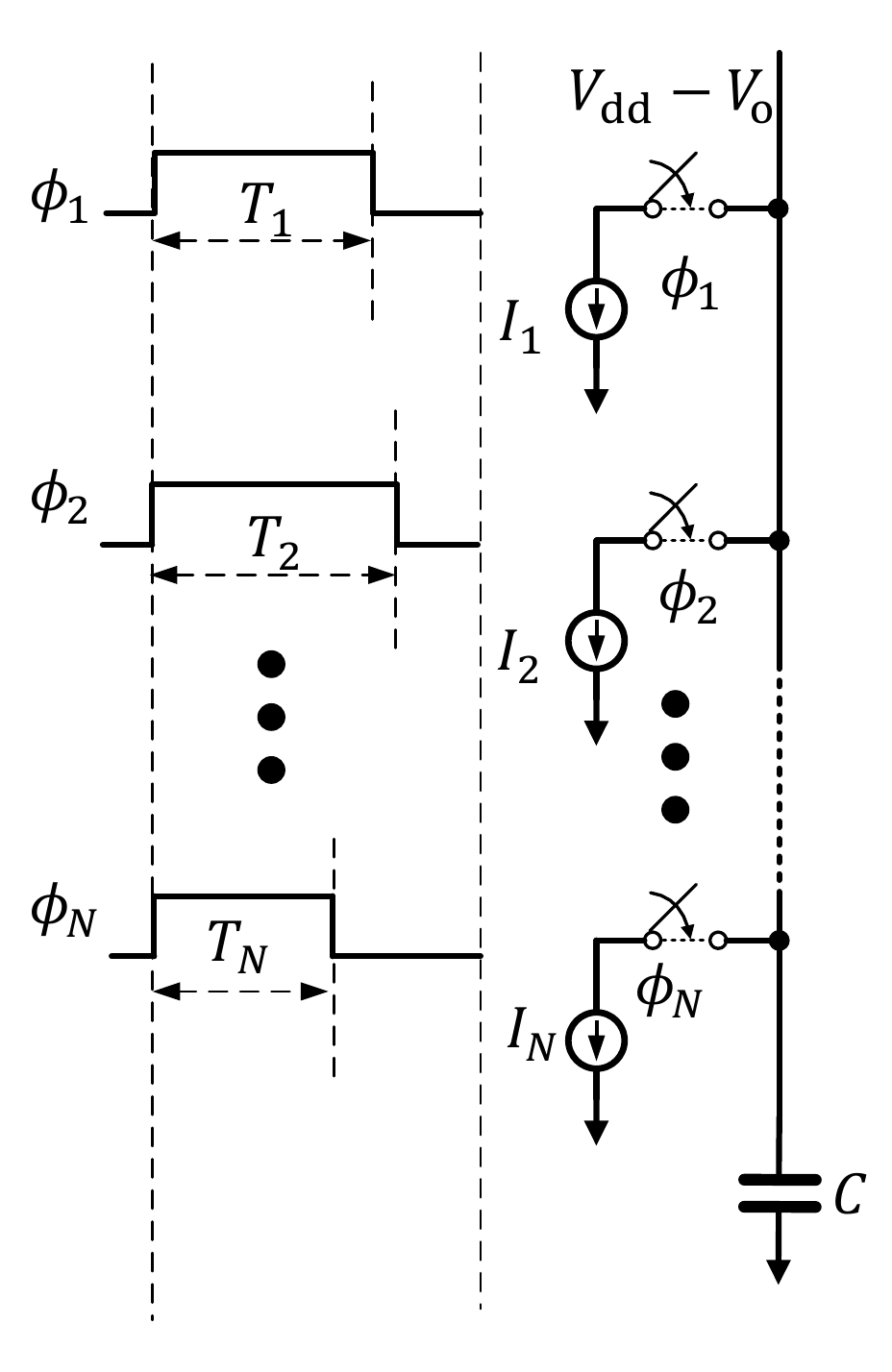}
    \label{fig:CharAcc}
    }$\quad$
    \subfloat[]{
    \includegraphics[width=0.2\linewidth]{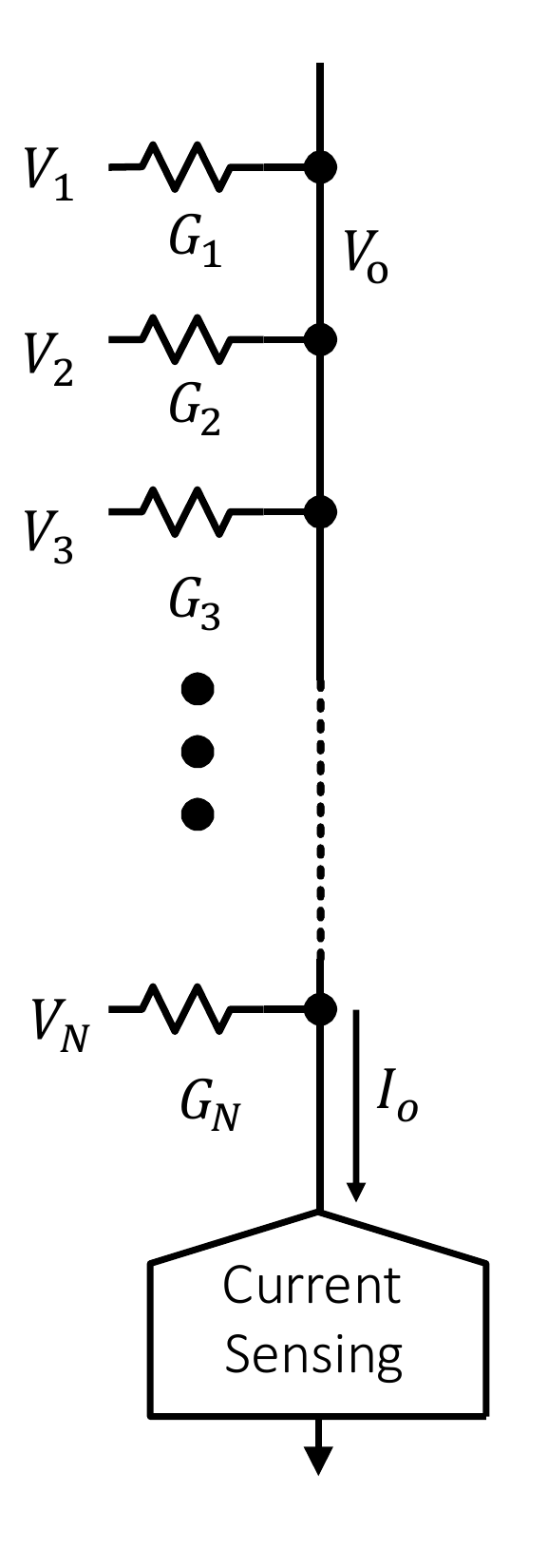}
    \label{fig:CurrSense}
    }
    
    \subfloat[]{
    \includegraphics[width=0.56\linewidth]{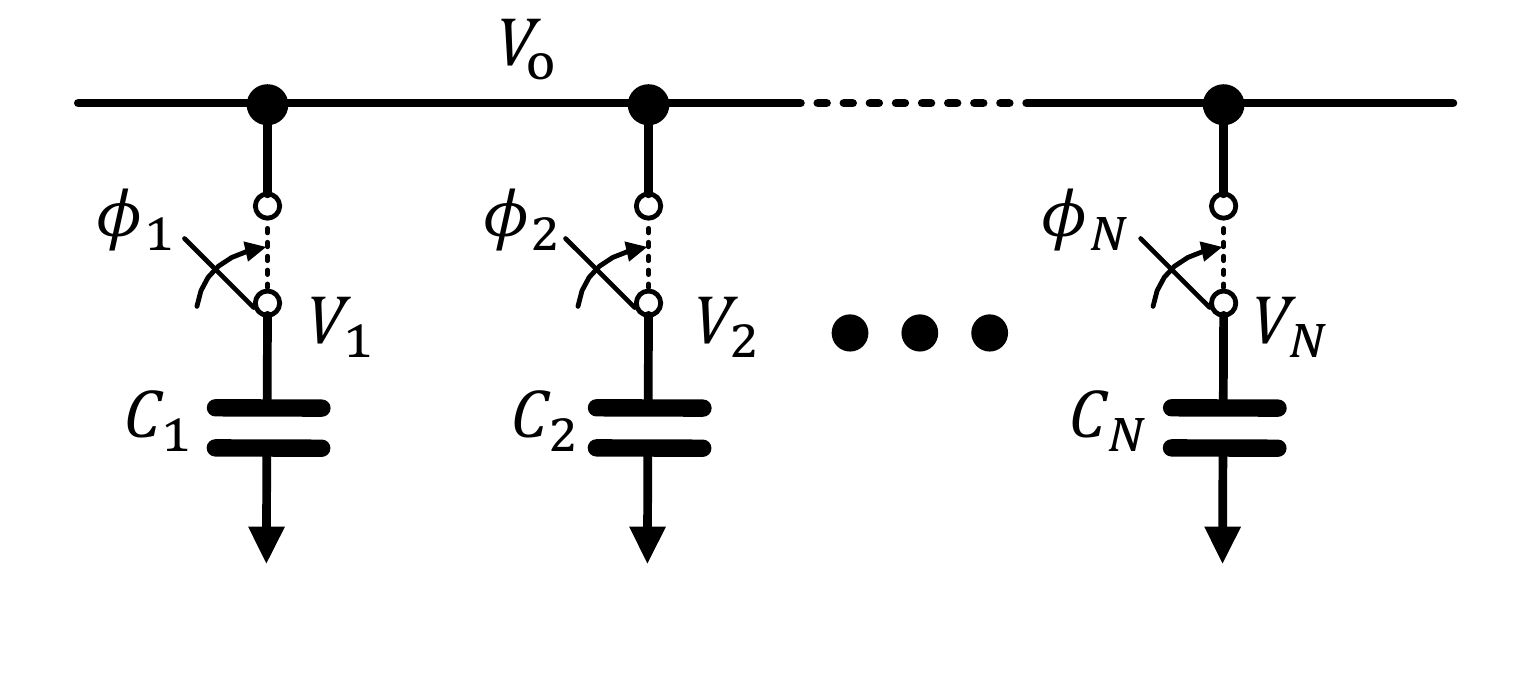}
    \label{fig:CharRe}
    }
    \caption{In-memory compute models: (a) charge summing (QS), (b) current summing (IS), and the (c) charge redistribution (QR) model.}
    \label{fig:compute-models}
\end{figure}

The QS model (see Fig.~\ref{fig:compute-models}(a)) realizes the DP in (\ref{eqn:DP}) via the variable mapping ($y_{\text{o}}\rightarrow V_{\text{o}}$, $w_j\rightarrow I_j$, $x_j\rightarrow T_j$) where the cell current $I_j$ is integrated over the WL pulse duration $T_j$ ($j=1,\ldots, N$) on a BL (or cell) capacitor $C$ resulting an output voltage as shown below:
\begin{align}
(y_{\text{o}}\rightarrow V_{\text{o}})&=\frac{1}{C}\sum_{j=1}^N (w_j \rightarrow I_j) (x_j \rightarrow T_j)
% V_{\text{o}}&=\frac{1}{C}\sum_{j=1}^N I_j T_j
    \label{eqn:QS-dot}
\end{align}
where $V_{\text{o}}$ is the DP output assuming infinite voltage head-room, i.e., no clipping. The cell current $I_j$ depends upon transistor sizes and the WL voltage $V_{\text{WL}}$, and 
%In other words, the QS model is defined by the mapping of variables in (\ref{eqn:DP}) as: $y_{\text{o}}\rightarrow V_{\text{o}}$, $w_j\rightarrow I_j$, and $x_j\rightarrow T_j$. 
typical values are: $C$ (a few hundred $\unit{fF}$s), $I_j$ (tens of $\unit{\mu A}$s), and $T_j$ (hundreds of $\unit{ps}$).

\begin{figure}
    \centering
    \subfloat[]{
    \includegraphics[width=0.2\linewidth]{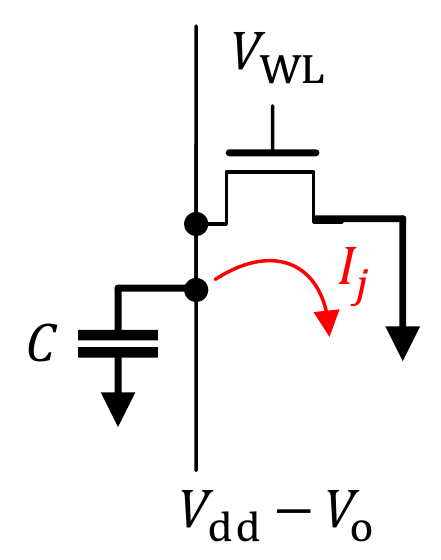}
    }
    \subfloat[]{
    \includegraphics[width=0.4\linewidth]{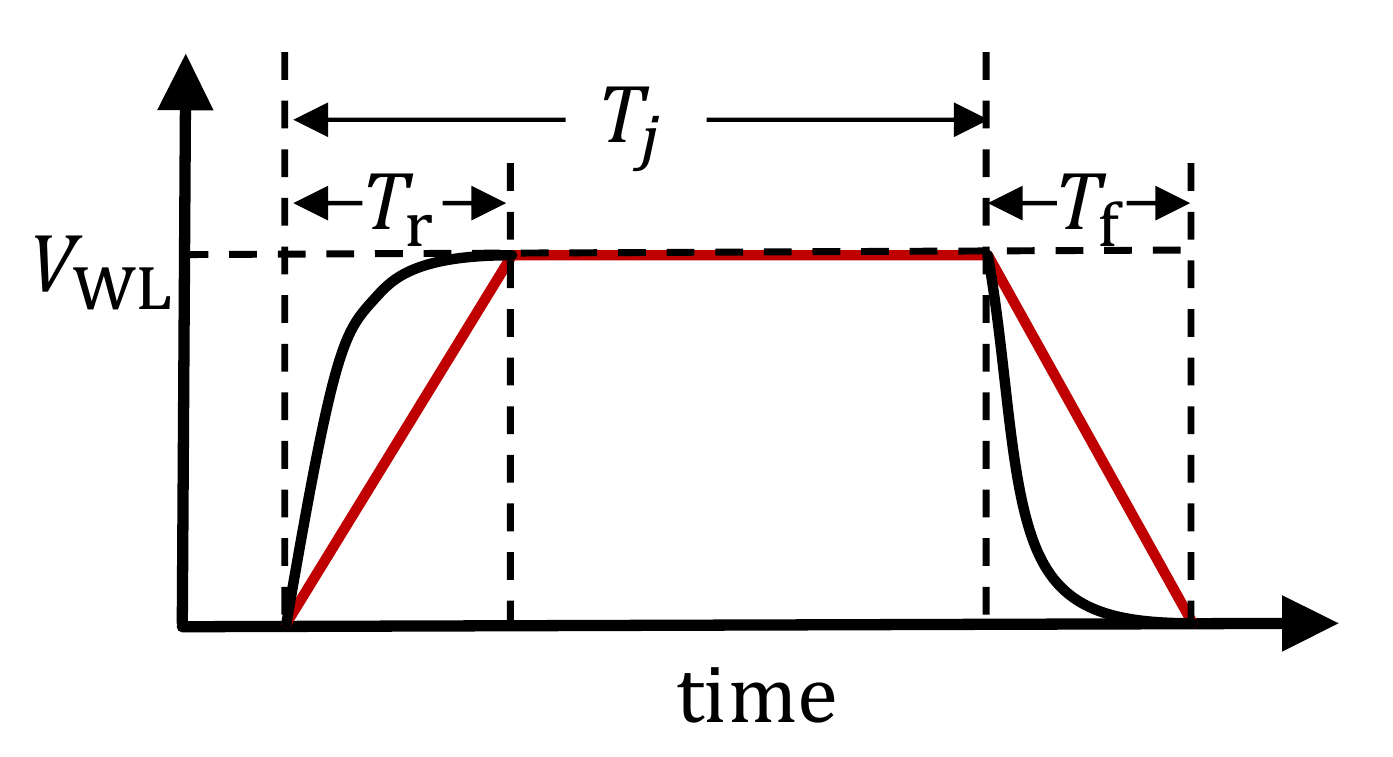}
    }
    \caption{Modeling the discharge process in the QS compute model: (a) cell current $I_j$, and (b) the word-line voltage pulse $V_{\text{WL}}$.}
    \label{fig:cell_pulse_model}
\end{figure}

\emph{Noise Models:}
The noise contributions in QS arise from the following sources: (1) variations in the pulse-widths $T_j$ of current switch pulses $\phi_j$ (Fig.~\ref{fig:compute-models}(a)); (2) their finite rise  and fall times (see Fig.~\ref{fig:cell_pulse_model}(b)); (3) spatial variations in the cell currents $I_j$; (4) thermal noise in the discharge RC-network; and (5) clipping due to limited voltage head-room. Thus, the analog DP output $V_{\text{a}}$ corresponding to $y_{\text{a}}=y_\text{o}+\etaa$ is given by:
\begin{align}
    (y_a \rightarrow V_\text{a}) &=(y_{\text{o}} \rightarrow V_{\text{o}}) +(\eta_{\text{e}} \rightarrow v_{\text{e}})+(\eta_\text{h} \rightarrow v_{\text{c}}),\nonumber\\
    v_{\text{e}}&=v_\theta+\frac{1}{C}\sum_{j=1}^N i_{j}T_j+I_j(t_{j}-t_{\text{rf}}),\nonumber\\
    v_{\text{c}}&=\min \left(V_{\text{o}},\ V_{\text{o,max}}\right)-V_{\text{o}},
    \label{eqn:QS-noise-model}
\end{align}
where $V_{\text{o,max}}$ is the maximum allowable output voltage, and $v_{\text{e}}$ and $v_{\text{c}}$ are the voltage domain noise due to circuit non-idealities and clipping, respectively, $i_{j}\sim\mathcal{N}(0,\sigma_{I_j}^2)$ is the noise 
due to (spatial) current mismatch, and $t_{j}\sim\mathcal{N}(0,\sigma_{T_j}^2)$ is the noise 
due to (temporal) pulse-width mismatch, respectively, both of which are modeled as zero mean Gaussian random variables, $t_{\text{rf}}$ models the impact of finite rise and fall times of the current switching pulses, and $v_\theta\sim\mathcal{N}(0,\sigma_{\theta})$ is the integrated thermal noise voltage. Note:  $V_{\text{o,max}}$ can be as high as \unit[0.9]{V} when $V_{\text{dd}}=\unit[1]{V}$.

Analytical expressions to estimate the noise standard deviations $\sigma_{I_j}$, $\sigma_{T_j}$, $\sigma_{\theta}$, and $t_{\text{rf}}$, (see appendix) are provided below:
\begin{align}
\sigma_{I_j} &=I_j\bigg(\frac{\alpha\sigma_{V_\text{t}}}{V_{\text{WL}}-V_\text{t}}\bigg)=I_j\sigma_{\text{D}}\label{eqn:current-mistmatch}\\
t_{\text{rf}} &= T_r-\Big(\frac{V_{\text{WL}}-V_\text{t}}{V_{\text{WL}}}\Big)\frac{T_{\text{r}} + T_{\text{f}}}{\alpha+1}\label{eqn:rise-fall-times}\\
\sigma_{T_j} &=\sqrt{h_j}\sigma_{T0},\quad
\sigma_\theta =\frac{1}{C}\sqrt{\frac{N T_\text{max}g_m kT}{3}}\label{eqn:qs-thermal}
\end{align}
where  $\sigma^2_{\text{D}}$ is normalized current mismatch variance, $T_j=h_jT_{\text{0}}$ is the delay of a $h_j$-stage WL driver composed unit elements with delay $T_{0}$ each, $\sigma_{T0}$ is the standard deviation of $T_0$,
$T_\text{r}$ and $T_\text{f}$ are WL pulse rise and fall times (see Fig.~\ref{fig:cell_pulse_model}(b)), $\alpha$ is a fitting parameter in the $\alpha$-law transistor equation,  $\sigma_{V_\text{t}}$ is standard deviation of $V_\text{t}$ variations, $k$ is the Boltzmann constant, $T$ is the absolute temperature, and $g_m$ is the transconductance of the access transistor.

Note that typically the WL voltage $V_{\text{WL}}$ is identical for all rows in the memory array with a few exceptions such as \cite{zhang_verma2017jssc} which modulate $V_{\text{WL}}$ to tune the cell current $I_j$. The effects of rise/fall times and delay variations can be mitigated by carefully designing the WL pulse generators. Therefore, noise in QS is dominated by spatial threshold voltage variations. Indeed, using the typical values from Table~\ref{tab:param}, we find that  $\sigma_{I_j}/I_j$ ranges from 8\% to 25\%, while $\sigma_{T_j}/T_j$ ranges from 0.5\% to 3\%. 

\begin{table}
\global\long\def\arraystretch{1.3}
    \centering
    \caption{In-Memory Compute Model Parameters in a representative $\unit[65]{nm}$ CMOS process}
    \begin{tabular}{|c||c|c||c|c|}
        %\hline
        %\multicolumn{4}{|c|}{Common Parameters}\tabularnewline
        \hline
         &Parameter & Value & Parameter & Value  \tabularnewline
         \hline
         \hline
          \multirow{4}{*}{QS}&$k'$ & \unit[220]{$\mu\text{A/V}^2$}  &  $\alpha$ & 1.8 \tabularnewline
         \cline{2-5}
          &$\sigma_{T0}$ & \unit[2.3]{ps} & $\sigma_{V_\text{t}}$ & \unit[23.8]{mV} \tabularnewline
         \cline{2-5}
        &$\Delta V_{\text{BL,max}}$  & $\unit[0.8]{V}$-$\unit[0.9]{V}$   & $V_{\text{WL}}$  &  $\unit[0.4]{V} - \unit[0.8]{V}$\tabularnewline
        \cline{2-5}
        %& $V_{\text{dd}}$  & $\unit[1]{V}$  & $C_{\text{BL}}$  & \unit[270]{fF}\tabularnewline
         %\cline{2-5}
         &$V_{\text{t}}$ & \unit[0.4]{V} & $T_{\text{0}}$ & \unit[100]{ps}\tabularnewline
        \hline
         \multirow{2}{*}{QR} &$WLC_\text{ox}$ & \unit[0.31]{fF} & $\kappa$  & \unit[0.08]{fF$^{0.5}$}\tabularnewline
         \cline{2-5}
         & $p$ & 0.5 & & \tabularnewline
        \hline 
        \multicolumn{5}{|c|}{Common Parameters}\tabularnewline
        \hline
        &$T$ & \unit[300]{K} & $k$ &$\unit[1.38\text{e-}23]{JK^{-1}}$\tabularnewline
        \cline{2-5}
        & $V_{\text{dd}}$ & \unit[1]{V} & $g_m$ & $\unit[66]{\mu A / V}$\tabularnewline
        \hline
    \end{tabular}
    \label{tab:param}
\end{table}

\emph{Energy and Delay Models:}
The average energy consumption in the QS model is given by:
\begin{align}
    E_{\text{QS}}=  \mathbb{E}\left[V_{\text{a}}\right] V_{\text{dd}} C +E_{\text{su}}
    \label{eqn:QS-energy}
\end{align}
where the spatio-temporal expectation $\mathbb{E}\left[V_{\text{a}}\right]$ is taken over inputs (temporal) and over columns (spatial)  $E_{\text{su}}$ is the energy cost of toggling switches $\phi_j$s. Equation (\ref{eqn:QS-energy}) shows that the energy consumption in the QS model increases with $C\propto$ array size, the supply voltage $V_{\text{dd}}$, and the mean value of the DP $\mathbb{E}\left[V_{\text{a}}\right]$.

The delay of the QS model is given by
$
    T_{\text{QS}}= T_{\max} +T_{\text{su}},
$
where $T_{\text{su}}$ is the time required to precharge the capacitors and setup currents, and $T_{\max}=\max\{T_j\}$ is the longest allowable pulse-width.

\subsubsection{The QS Architecture (\QSArch{})}

\begin{figure*}
    \subfloat[]{
    \centering
    \includegraphics[height=1.8in]{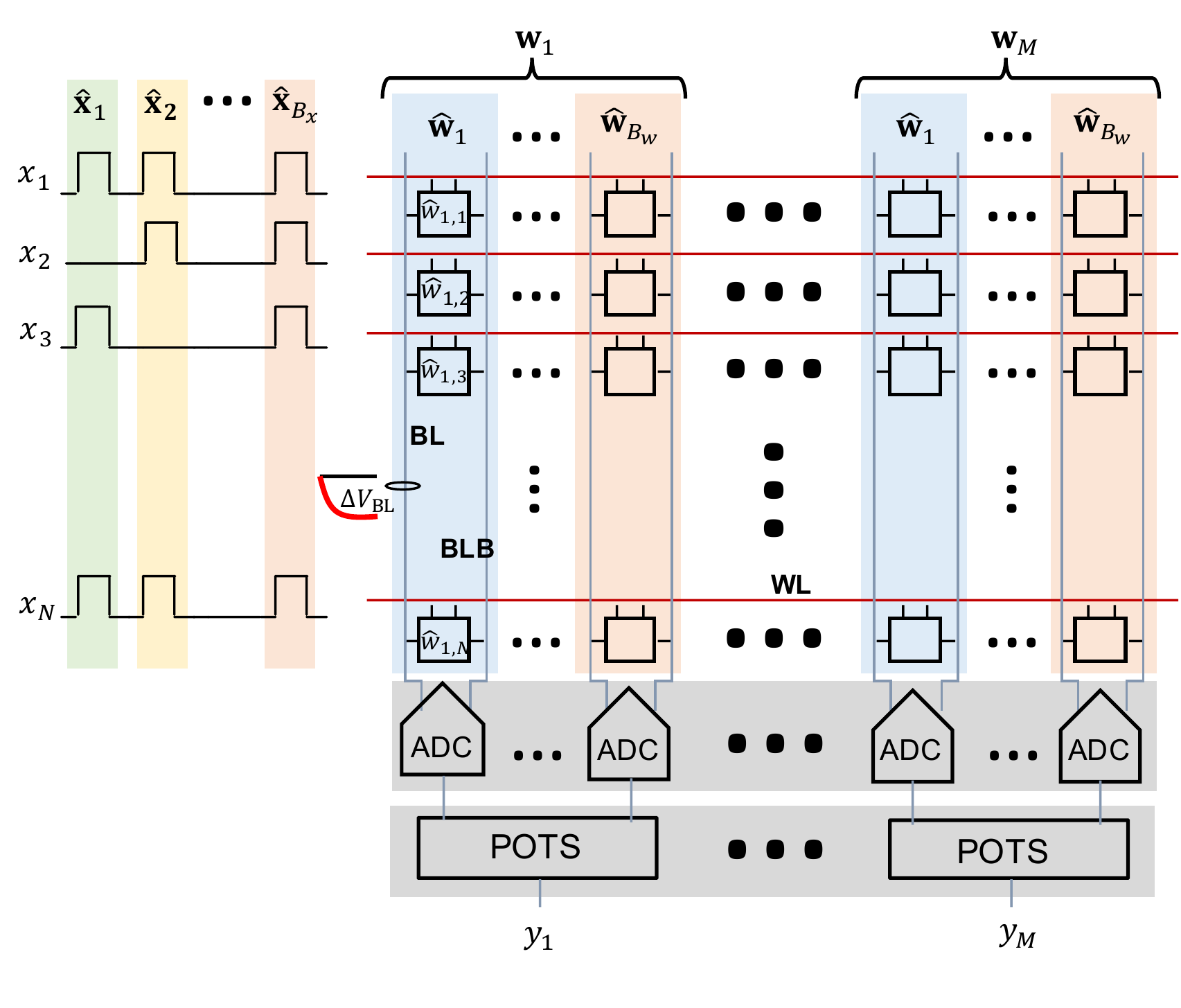}}
    \subfloat[]{
    \centering
    \raisebox{0.2in}{\includegraphics[height=1.4in]{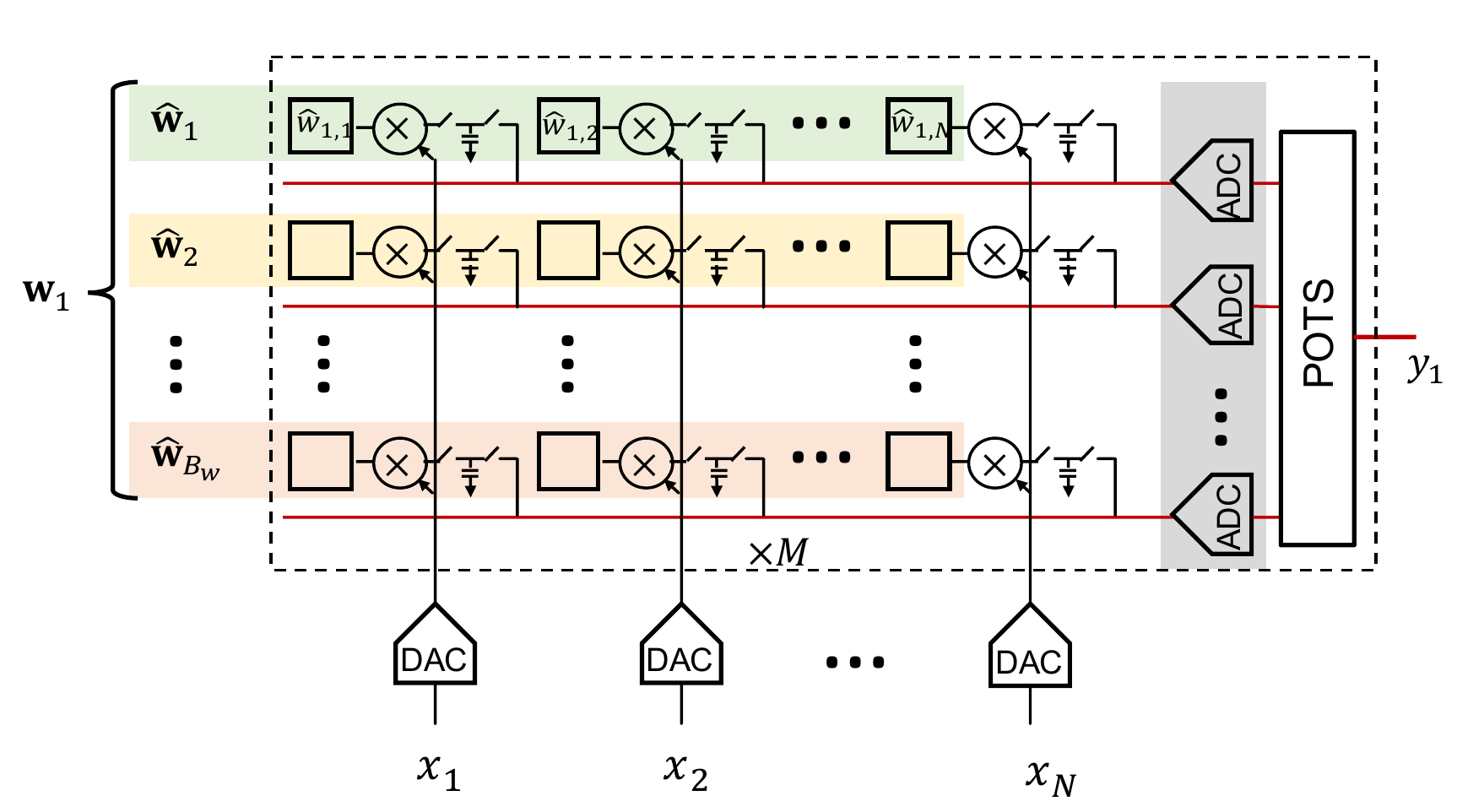}}}
    \subfloat[]{
    \centering
    \includegraphics[height=1.8in]{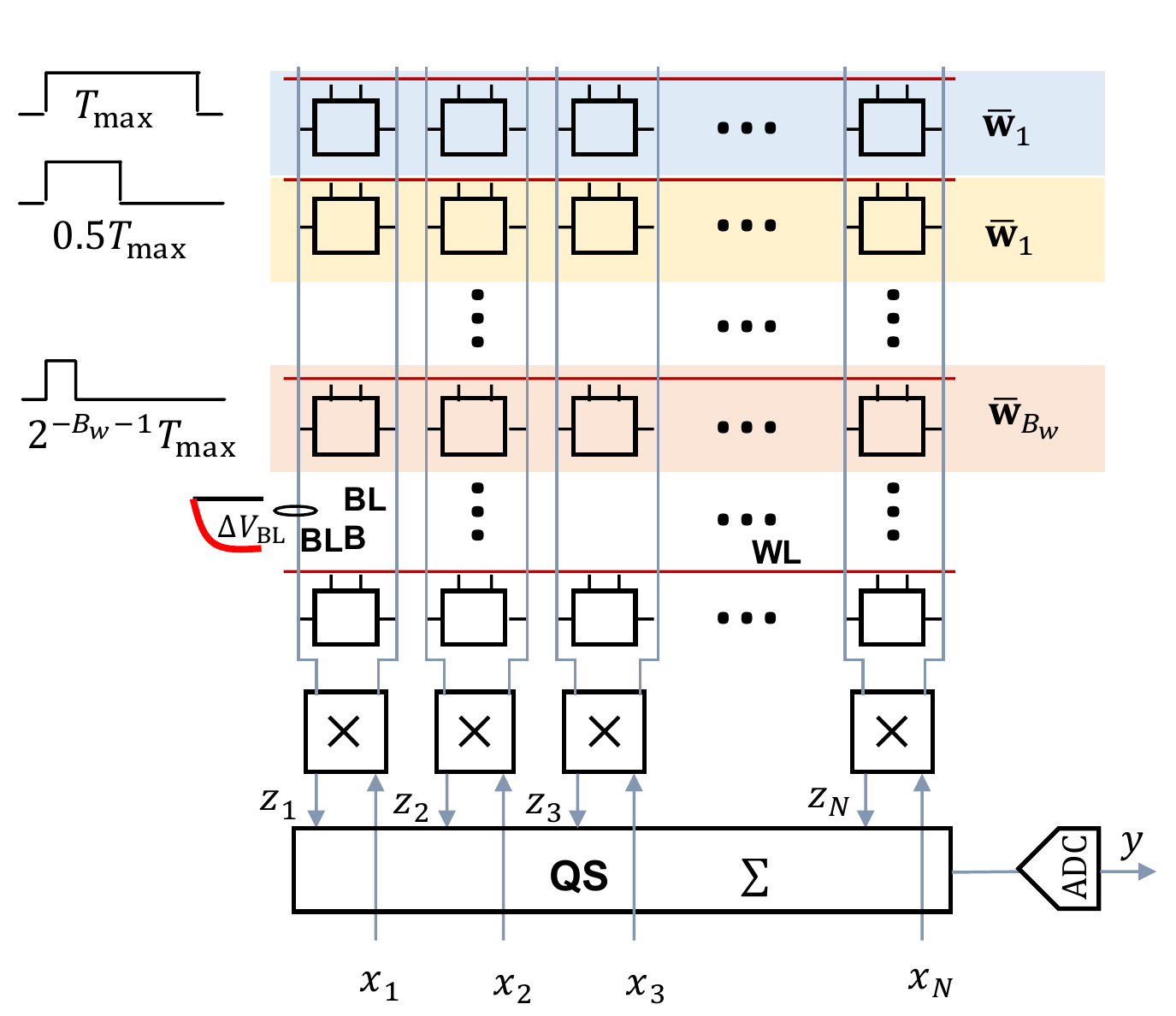}}
    \centering
    
    \caption{Mapping of multi-bit DPs on in-memory architectures: (a) \QSArch{}, (b)  QR-Arch, and (c) CM.}
    \label{fig:Architectures}
\end{figure*}

The charge summing architecture (\QSArch{}) in Fig.~\ref{fig:Architectures}(b) employs a 6T \cite{guo2019vlsi} or 8T \cite{si2019isscc} SRAM bitcell 
within the QS model (see Section~\ref{subsec:QS-model}).
This architecture implements fully-binarized DPs on the BLs by mapping the input bit $\hat{x}_{i,j}$ to the WL access pulse $V_{\text{WL},j}$ while the weights $\hat{w}_{i,j}$ are stored across $B_w$ columns of the BCA so that the BC currents $I_{i,j}\propto \hat{w}_{i,j}$. The output $V_{\text{o}}=\Delta V_{\text{BL}}$ is the voltage discharge on the BL and the capacitance $C=C_{\text{BL}}$ is the BL capacitance in (\ref{eqn:QS-dot}). \QSArch{} sequentially (bit-serially) processes one multi-bit input vector $\mathbf{x}$ in $B_x$ in-memory compute cycles followed by a digital summing of the binarized DPs to obtain the final multi-bit DP (\ref{eqn:DP}).

\begin{table*}[]
    \global\long\def\arraystretch{1.5}
    \caption{Derived noise and precision parameters for \QSArch{}, \QRArch{} and CM using the compositional framework.}
    \label{tab:arch_summary}
    \centering
    
	\resizebox{0.98\linewidth}{!}{
		\centering
    {\tabulinesep=1.2mm
    \begin{tabu}{|c||c|c|c|}
            %\hline
            %& $\sigma^2_\text{q}$ & $\sigma^2_\text{c}$ & $\sigma^2_\text{e}$ 
            %\tabularnewline
            \hline
          & \textbf{\QSArch{}} & \textbf{\QRArch{}}& \textbf{CM}

        \tabularnewline
        \hline
        \hline
        Bitcell type
        
        &
        6T or 8T
        &
        8T or 10T + MOM cap &
        6T

        \tabularnewline
        \hline
        \parbox[c]{1.5cm}{\centering Analog Core Precision}
   
        &
        Binarized ($B_w=B_x=1$)
        &
        Binary-weighted ($B_w=1$)     &
        Multi-bit

        \tabularnewline
        \hline
        \parbox[c]{1.5cm}{\centering Compute  model  used}
      
        &
        QS
        &
        QR
    &
        QS \& QR
        
        \tabularnewline
        \hline
        \parbox[c]{1.5cm}{\centering Energy cost per DP}
   
        &
        $E_{\text{\QSArch{}}}=B_wB_x(E_{\text{QS}}+E_\text{ADC}) +E_{\text{misc}}$
        &
        $\begin{array}{lcl}E_{\text{\QRArch{}}}=B_w(E_{\text{QR}}+NE_{\text{mult}}+E_\text{ADC})+E_{\text{misc}} \\ E_{\text{mult}}=\mathbb{E}[x_j(1-w_j)]C_{\text{o}}V_{\text{dd}}]\end{array}$ 
     &
        $\begin{array} {lcl} E_{\text{CM}}=2NE_{\text{QS}}+E_{\text{QR}} \\ +E_{\text{mult}}+E_\text{ADC}+E_{\text{misc}} \end{array}$
        
        \tabularnewline
        \hline
        \parbox[c]{1.5cm}{\centering Compute model mapping}
        &
        \parbox[c]{2.3cm}{  $C \rightarrow C_{\text{BL}}$ \\ $V_{\text{o}} \rightarrow \Delta V_{\text{BL}}$ \\ $T_{j} \rightarrow T_{\text{WL}, j} $ }
        &
        $C_j \rightarrow C_{\text{o}}$
        &
        \parbox[c]{2.3cm}{ QS:  $C \rightarrow C_{\text{BL}}$ \\ $V_{\text{o}} \rightarrow \Delta V_{\text{BL}}$ \\ $T_{j} \rightarrow T_{\text{WL}, j} $ \\ QR: $C_j \rightarrow C_{\text{o}}$}

        \tabularnewline
        \hline
         $\sigma^2_{q_{iy}}$ 
  
         &  
         $\frac{1}{12}N\Delta_x^2\sigma^2_w + \frac{1}{12}N\Delta_w^2\mathbb{E} \left[ x^2 \right]$
         &
         $\frac{1}{12}N\Delta_x^2\sigma^2_w + \frac{1}{12}N\Delta_w^2\mathbb{E} \left[ x^2 \right]$
       &
         $\frac{1}{12}N\Delta_x^2\sigma^2_w + \frac{1}{12}N\Delta_w^2\mathbb{E} \left[ x^2 \right]$

            \tabularnewline
            \hline       
       $\sigma^2_{\eta_{\text{h}}}$       
         & 
        
        $ \begin{array} {lcl} \frac{4}{9}\left(1-4^{-B_w}\right)\left(1-4^{-B_x}\right)
        \\
        \sum_{k=k_{\text{h}}}^N \left(k-k_{\text{h}}\right)^2\binom{N}{k} \left(\frac{1}{4}\right)^k\left(\frac{3}{4}\right)^{N-k} \end{array}$ 
        
        &
        
        0
        & $ \begin{array} {lcl} \frac{1}{12}N\mathbb{E} \left[ x^2 \right] \sigma^2_wk_{\text{h}}^{-2}2^{2B_w} \\ \left(1-2k_{\text{h}}2^{-B_w}\right)_+^2 \end{array}$

        \tabularnewline
        \hline         
         
        $\sigma^2_{\eta_{\text{e}}}$ 
         
         &
         $\frac{N\sigma_\text{D}^2\left(1-4^{-B_w}\right)\left(1-4^{-B_x}\right)}{9}$ 
         
         &
         $\begin{array}{lcl} \frac{2}{3}(1-4^{-B_w}) N 
\bigg(\frac{\mathbb{E} \left[ x^2 \right]\sigma_{C_\text{o}}^2}{C_{\text{o}}^2}
+\frac{2\sigma_{\theta}^2}{V_\text{dd}^2}+\sigma_{\text{inj}}^2 \bigg) \end{array}
$&
        
        $\frac{2}{3}N\mathbb{E} \left[ x^2 \right]\left(\frac14-4^{-B_w}\right)\sigma_\text{D}^2$ 
         
        \tabularnewline
        \hline

        $B_\text{ADC}$  
        & 
        $\geq \min \big(\frac{\SNRAdb+16.2}{6},\log_2 (k_{\text{h}}), \log_2 (N) \big) $ 
        & 
        $\geq \min \big( \frac{\SNRAdb+16.2}{6}, B_x + \log_2 (N) \big)$
        
        & 
        $\geq \frac{\SNRAdb+16.2}{6}$

        \tabularnewline
        \hline
        
        $V_{\text{c}}$ 
        &
        $\min \bigg( 4\sqrt{3N}  \Delta V_{\text{BL},\text{unit}}  , \Delta V_{\text{BL,\text{max}}}, N \Delta V_{\text{BL},\text{unit}}  \bigg) $
        &
        $ 8V_{\text{dd}}\sqrt{\frac{\mathbb{E} \left[ x^2 \right]+\sigma_x^2}{N}}$
& $\frac{8\sigma_w 2^{B_w}\Delta V_{\text{BL},\text{unit}} \sqrt{\mathbb{E}\left[ x^2 \right]}}{\sqrt{N}}$ 
        \tabularnewline
        \hline        
        
    \end{tabu}
    }
    }
    
    \vspace{4pt}
$k_{\text{h}} = \frac{\Delta V_{\text{BL.max}}}{\Delta V_{\text{BL,unit}}}$; $\sigma_\text{D}=\frac{\sigma_{I}}{I}$ is the normalized standard deviation of the bit-cell current (\ref{eqn:current-mistmatch}); $(x)_+=\max(x,0)$; $\sigma^2_{\text{inj}}=\mathbb{E}\left[x^2\right] WLC_{\text{OX}}/C_{\text{o}}$ .

\end{table*}

We derive the analytical expressions of architecture-level noise models for \QSArch{} using those of the QS model described in Section~\ref{subsec:QS-model}. 
In \QSArch{}, clipping occurs in each of the $B_x\times B_w$ binarized DPs and contributes to the overall clipping noise variance $\sigma^2_{\eta_{\text{h}}}$ at the multi-bit DP output. Circuit noise from each binarized DP is aggregated to obtain the final circuit noise variance $\sigma^2_e$. In addition, employing MPC imposed requirement on the final DP output precision $B_y$ \eqref{eqn:bympc}, we obtain the lower bound on ADC precision $B_{\text{ADC}}$.

%Table~\ref{tab:param} tabulates the parameters of all three in-memory compute models in a representative \unit[65]{nm} CMOS process. These models describe the realization of a multi-bit DP (\ref{eqn:DP}) by mapping the inputs $x_j$, weights $w_j$ and the output $\yo$ to appropriate physical variables. 
Since the multi-bit DP computation in (\ref{eqn:DP}) is high-dimensional ($N$ can be in hundreds), it is clear that the limited BL dynamic range e.g., $V_{\text{o,max}}$ in \eqref{eqn:QS-noise-model}, will begin to dominate $\SNRa$ in \eqref{eqn:imc-accuracy-metrics}. It is for this reason that most, if not all, IMCs resort to some form of binarization of the multi-bit DP in (\ref{eqn:DP}) prior to employing one of the in-memory compute models (see Table~\ref{tab:classify}). Ultimately, $\SNRa$ limits the number and accuracy of BL computations per read cycle and hence the overall energy efficiency of IMCs.

\color{black}
\subsection{Charge Redistribution (QR)}\label{subsec:QR-model}
\subsubsection{The QR model}
The QR model (Fig.~\ref{fig:compute-models}(c)) is commonly employed to perform the additions in \eqref{eqn:DP}. The multiplications in \eqref{eqn:DP} are separately computed via charging/discharging capacitor $C_j$ in proportion to the product $w_jx_j$ ($j=1,\ldots,N$) as in \cite{valavi2018vlsi,biswas2018conv}, or by employing explicit multiplier circuits such as in \cite{kang2018jssc,gonugondla2018jssc}. The $N$ capacitors share charge via a sequence of switching events (see Fig.~\ref{fig:compute-models}\subref{fig:CharRe}) to generate the final voltage $V_{\text{o}}$ given by:
\begin{align}
    (y_{\text{o}} \rightarrow V_{\text{o}}) = \frac{1}{ \sum_{j} C_{j}} \sum_{j}C_{j} (w_j x_j \rightarrow V_j)\label{eqn:QR-model}
\end{align}
The capacitors $C_j$ are typically metal-on-metal (MOM) capacitors with values ranging from \unit[1]{fF} to \unit[10]{fF} \cite{kang2018jssc,valavi2018vlsi}.

\emph{Noise Models:}
Assuming MOM-based $C_j$s, the noise contributions in QR arise from: (1) capacitor mismatch \cite{tripathimismatch}; (2) charge injection due to switching \cite{chargeinjection}; and (3) thermal noise.  Unlike QS, and similar to IA, the QR model does not suffer from headroom clipping noise. Hence, the DP output $V_{\text{a}}$  corresponding to $y_{\text{a}}$ in (\ref{eqn:imc-noise-model}) is given by:
\begin{align}
    (y_{\text{a}} \rightarrow V_{\text{a}})&=(y_{\text{o}} \rightarrow V_{\text{o}}) + (\eta_{\text{e}} \rightarrow v_{\text{e}}) \label{eqn:QR-noise-model}\\
    &= \frac{1}{\sum_{j} (C_{j} +c_{j} )} \sum_{j} (C_{j} +c_{j} )(V_j +v_{\theta_j} +v_{j}) 
    \nonumber
\end{align}
where $v_{\text{e}}$ is the voltage domain noise term due to circuit non-idealities corresponding to $\eta_{\text{e}}$ in (\ref{eqn:imc-noise-model}), $v_{j}$ is the noise is due to charge injection, $c_{j}\sim \mathcal{N}(0,\sigma_{C_j}^2)$ is the capacitor mismatch, and $v_{\theta_j}\sim \mathcal{N}(0,\sigma_{\theta,j})$ is the thermal noise. Furthermore, expressions for the noise parameters in (\ref{eqn:QR-noise-model}) can be derived as \cite{tripathimismatch,chargeinjection}:
\begin{align}
    \sigma_{C_j}=\kappa \sqrt{C_{j}},~ %\label{eqn:capmis}\\
    v_{j} = p\frac{WLC_\text{ox} (V_{\text{dd}}-V_\text{t} -V_{j})}{C_{j}},~%\label{eqn:chinj}\\
    \sigma_{\theta,j}=\sqrt{\frac{kT}{C_{j}}}\label{eqn:chinj}%\label{eqn:qs-thermal}
\end{align}
where $\kappa$ is a technology- and layout-dependent Pelgrom coefficient \cite{tripathimismatch}, $0\leq p\leq 1$ is constant that depends on the layout of the switch transistor, $C_\text{ox}$ is the gate oxide capacitance per unit area, and $W$ and $L$ are the width and length of the switch transistor. The effect of noise in the QR compute model can be minimized by increasing the capacitors sizes at the expense of energy consumption as seen from \eqref{eqn:QR-energy} below.

\emph{Energy and Delay Models:}
The average energy consumption in the QR model is given by:
\begin{align}
    E_{\text{QR}}= \sum_j \mathbb{E}\left[(V_{\text{dd}}-V_j)\right]V_{\text{dd}}C_{j} +E_{\text{su}} 
    \label{eqn:QR-energy}
\end{align}
where $E_{\text{su}}$ includes energy cost for the switches $\phi_j$s.   

The delay of the QR model is given by:
$
    T_{\text{QR}}= T_{\text{share}} +T_{\text{su}},
$ 
where $T_{\text{share}}$ is the time required for charge sharing to complete, and $T_{\text{su}}$ is the time required to precharge the capacitors to the desired voltages $V_j$.

%\subsection{Summary of In-memory Compute Models}
\subsubsection{The QR architecture (\QRArch{})}

The \QRArch, e.g., \cite{biswas2018conv},  in Fig.~\ref{fig:Architectures}(c) employs a modified BC to include a capacitor $C_{\text{o}}$ and additional switches for multiplication within the QR model. While works such as \cite{biswas2018conv} employ the parasitic capacitances on the BL within the BC, an explicit MOM capacitor is assumed for simplicity. \QRArch{} implements a binary weighted DP by storing the weights $\hat{w}_{i,j}$ across $B_w$ rows of the BCA and by providing a multi-bit analog input $x_{j}$ to the multiplier. The multiplication is implemented by first charging the capacitor $C_{\text{o}}$ to a voltage proportional to ${x}_{j}$ and then discharging it based on  $\hat{w}_{i,k}$. Multiplication is followed by a QR operation across the rows so that the final voltage across the capacitors in each row is proportional to binary-weighted DP. Thus, the \QRArch{} processes one multi-bit input vector $\mathbf{x}$ in one in-memory compute cycle to compute binary-weighted DPs that are power-of-two (POT) summed digitally to obtain the final multi-bit DP (\ref{eqn:DP}).

The average energy per DP $E_\text{\QRArch{}}$ (see Table~\ref{tab:arch_summary}) includes
$E_{\text{QR}}$ (\ref{eqn:QR-energy}) and $E_{\text{misc}}$ which in turn includes the energy consumption of the DACs used for converting $x_j$ into the analog domain. The energy cost of DACs are amortized since these are shared by multiple DPs computations.

Since the QR compute model does not suffer from headroom clipping, $\sigma_\text{h}^2=0$ and the lower bound on $B_{\text{ADC}}$ is dictated solely by MPC. The primary analog noise contributors ($\sigma^2_{\eta_{\text{e}}}$) are the capacitor mismatch ($\sigma^2_{C_\text{o}}$), charge injection noise ($\sigma^2_{\text{inj}}$) in the switches, and thermal noise ($\sigma^2_{\theta}$) as indicated in Table~\ref{tab:arch_summary}. An alternative QR model \cite{bankman20168,jiang2020c3sram,murmann2020tvlsi} is to directly switch the bottom plate of the capacitors with binary outputs of the BC multiplies. Doing so leads to greater energy efficiencies and significant reduction in charge injection noise. The analysis presented in this section can be extended to such models as well.

%the primary source of noise is due to circuit non-idealities ($\sigma^2_{\eta_{\text{e}}}$). Circuit noise variance $\sigma^2_{\eta_{\text{e}}}$ shown in Table~\ref{tab:arch_summary} is derived from (\ref{eqn:QR-noise-model}), and (\ref{eqn:chinj}) as shown in the appendix. Standard deviation of charge injection noise $\sigma_{\text{Inj}}=\sqrt{N \mathbb{E}\left[x^2\right]}\sigma_w WLC_{\text{OX}}/C_{\text{o}}$ is derived from \eqref{eqn:chinj}, where we assume the mean of charge injection noise does not affect the SNR as it can be easily corrected for. 

%\QRArch{} invokes $B_w$ ADCs of precision $B_y$ per DP. Since t maintain $\text{SNR}_{\text{d}}>\text{SNR}_{\text{a}}+\unit[9]{dB}$, we need a precision $B_y$ using MPC (\ref{eqn:sqnr-mpc})
%and the corresponding quantization range in the voltage domain ($V_{\text{c}}$) are listed in Table~\ref{tab:arch_summary}.
%Figure~\ref{fig:BM-ChSh-SNR}(b) shows that predicted $B_y$ based on MPC with $\zeta_y^{\text{MPC}}$ results in a SNR loss of less than $\unit[0.5]{dB}$ as intended.

\subsection{Compute Memory}

CM \cite{kang2014icassp,kang2018jssc,gonugondla2018jssc} in Fig.~\ref{fig:Architectures}(c) employs a 6T SRAM BC within the QS (see Section~\ref{subsec:QS-model}) and QR (see Section~\ref{subsec:QR-model})  models. In the most general case, CM strives to implement a $B_w\times B_x$-b DP directly by mapping $B_x$-b inputs $x_j$ to pulse width $T_j$ and/or amplitude $V_{\text{WL},j}$ of the WL access pulses and storing $B_w$-b weights $w_j$ in a column-major format across $B_w\times N$ cells. In practice, CM realizations such as \cite{kang2018jssc} employ POT weighted WL access pulse-widths for $B_w$ rows so that the voltage discharge $\Delta V_{\text{BL}}$ on the $j$-th BL is proportional to the weight $w_j$. The product $w_jx_j$ is realized using a per-column charge redistribution-based multiplier, followed by a QR stage to aggregate the $N$ multiplier outputs. In this way, CM computes the $B_w\times B_x$-b DP (\ref{eqn:DP}) in analog in a single in-memory compute cycle.

The energy cost per DP for CM in Table~\ref{tab:arch_summary} is obtained by substituting $C=C_{\text{BL}}$ and $V_{\text{o}}=\Delta V_{\text{BL}}$ in (\ref{eqn:QS-energy}), and using (\ref{eqn:QR-energy}) with $C_j=C_{\text{o}}$. Here, $E_{\text{mult}}$ is the energy consumption of the mixed-signal multiplier. The factor of 2 in first term accounts for discharge on both BL and BL-bar needed to realize signed weights \cite{gonugondla2018jssc}. The second term is the energy consumed in aggregating the column outputs using the QR model with identical capacitors $C_{\text{o}}$. 

%The expressions  $\sigma^2_{\eta_{\text{h}}}$ and $\sigma^2_{\eta_{\text{e}}}$ for CM can be derived from the noise models of the compute model (see Appendix). The derived expressions are shown in Table~\ref{tab:arch_summary}, 

The expression for $\sigma^2_{\eta_{\text{e}}}$ neglects the impact of pulse width variations and other noise sources in QR since it is dominated by variations in the bit-cell current $I_j$. However, the sample-accurate Monte Carlo simulations incorporte all noise sources.

%The ADC needs to quantize the output $y$ in the range $[-\zeta \sigma_y, \zeta \sigma_y]$. In order to design the appropriate ADC for CM we need to  find the voltage domain equivalent for this quantization range. The quantization range in the voltage domain ($V_{\text{c}}$) at the ADC input in CM (see Fig~\ref{fig:Architectures}(a)) is shown in Table~\ref{tab:arch_summary},
%where $\Delta V_{\text{BL},\text{unit}}$ is the unit bitline discharge. 
The factor of $\sqrt{N}$ in the denominator of the expression for the ADC input range $V_{\text{c}}$ due to the use of QR indicates that the loss in voltage range due to charge redistribution across $N$ capacitors. 
%The corresponding ADC quantization step ($\Delta_\text{ADC}$) in voltage domain is given by $V_{\text{c}}/2^{B_y}$. 

\color{black}

\section{Simulation Results}

\label{sec:simulations}
This section describes the SNR validation methodology for validating the SNR (noise) expressions in Table~\ref{tab:arch_summary} and simulation results for \QSArch{}.
%The noise models summarized in Table~\ref{tab:arch_summary} can be used by circuit designers and architects to evaluate the accuracy vs. energy trade-offs within the its design space occupied by IMC architectures without requiring computationally-intensive circuit simulations.  
\subsection{SNR Validation Methodology}
\label{sec:validation}
\begin{figure}
    \centering
    \includegraphics[width=0.9\linewidth]{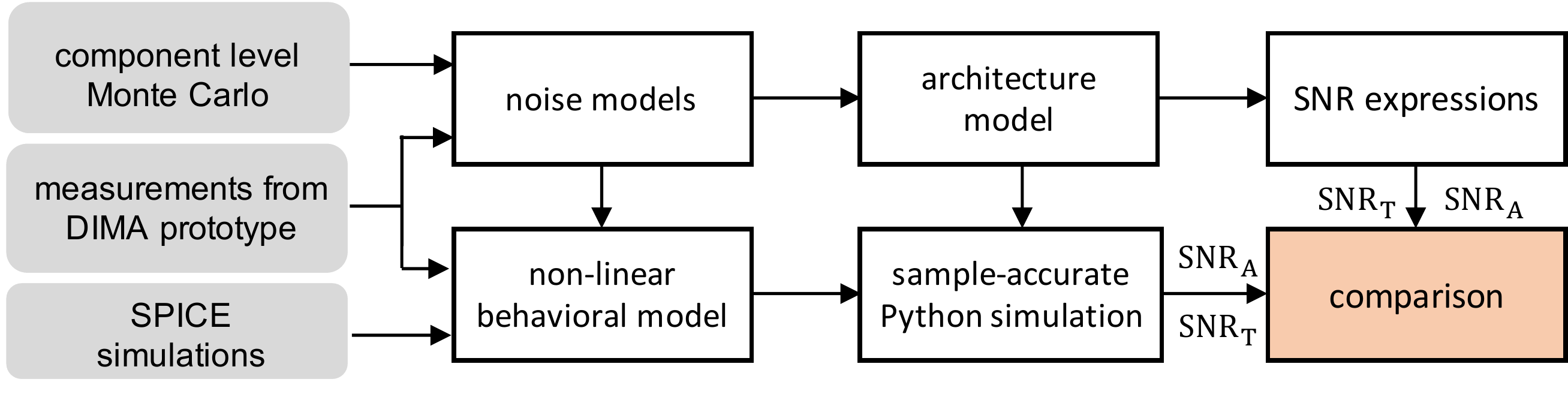}
    \caption{SNR validation methodology.}
    \label{fig:methodology}
\end{figure}

 Figure~\ref{fig:methodology} describes the SNR validation methodology. We obtain the QS and QR model parameters  (Section~\ref{sec:compute-models}) using Monte Carlo circuit simulations in a representative \unit[65]{nm} CMOS process, with experimental validation of some of these, e.g., $\sigma_{\eta_\text{e}}$, from our IMC prototype ICs \cite{gonugondla2018isscc,kang2018jssc} when possible. %The final set of compute model parameters are summarized in Table~\ref{tab:param}. 

Incorporating non-linear circuit behavior along with noise models, sample-accurate Monte Carlo Python simulations are employed to numerically calculate $\text{SNR}$ values using ensemble averaged (over 1000 instances) statistics. We compare the $\text{SNR}$ values obtained through sample-accurate simulations with those obtained by evaluating the analytical expressions in Table~\ref{tab:arch_summary}.

The quantitative results in subsequent sections employ the QS and QR model parameter values in Table~\ref{tab:param} along with \QSArch{}, \QRArch{}, and CM energy and noise models from Table~\ref{tab:arch_summary}.
An SRAM BCA with 512 rows and $C_{\text{BL}}=\unit[270]{fF}$ is assumed throughout. 
%We assume the impact of rise and fall time $\gamma_{rf}$ (\ref{eqn:rise-fall-times}) can be minimized via calibration.
\color{black} 
Energy and accuracy is traded-off of by tuning $V_{\text{WL}}$ in \QSArch{} and CM, and by tuning $C_\text{o}$ in \QRArch{} We assume zero mean signed weights $w_j$ and unsigned inputs $x_j$ drawn independently from two different distributions.\color{black} We set $B_x=B_w=6$ everywhere, unless otherwise stated, so that $\SQNRqiydb=\unit[38.9]{dB}\gg\SNRadb$ and therefore $\SNRA\approx\SNRa$ from \eqref{eqn:SNRA}. Next, we show how $\SNRA$ and $\SNRT$ trade-off with $N$ and $B_\text{ADC}$.

\subsection{SNR Trade-offs in IMCs}

\begin{figure*}[!t]
    \begin{minipage}{0.3\textwidth}\vspace{0pt}
        \centering
        \subfloat[]{
        \includegraphics[width=5cm]{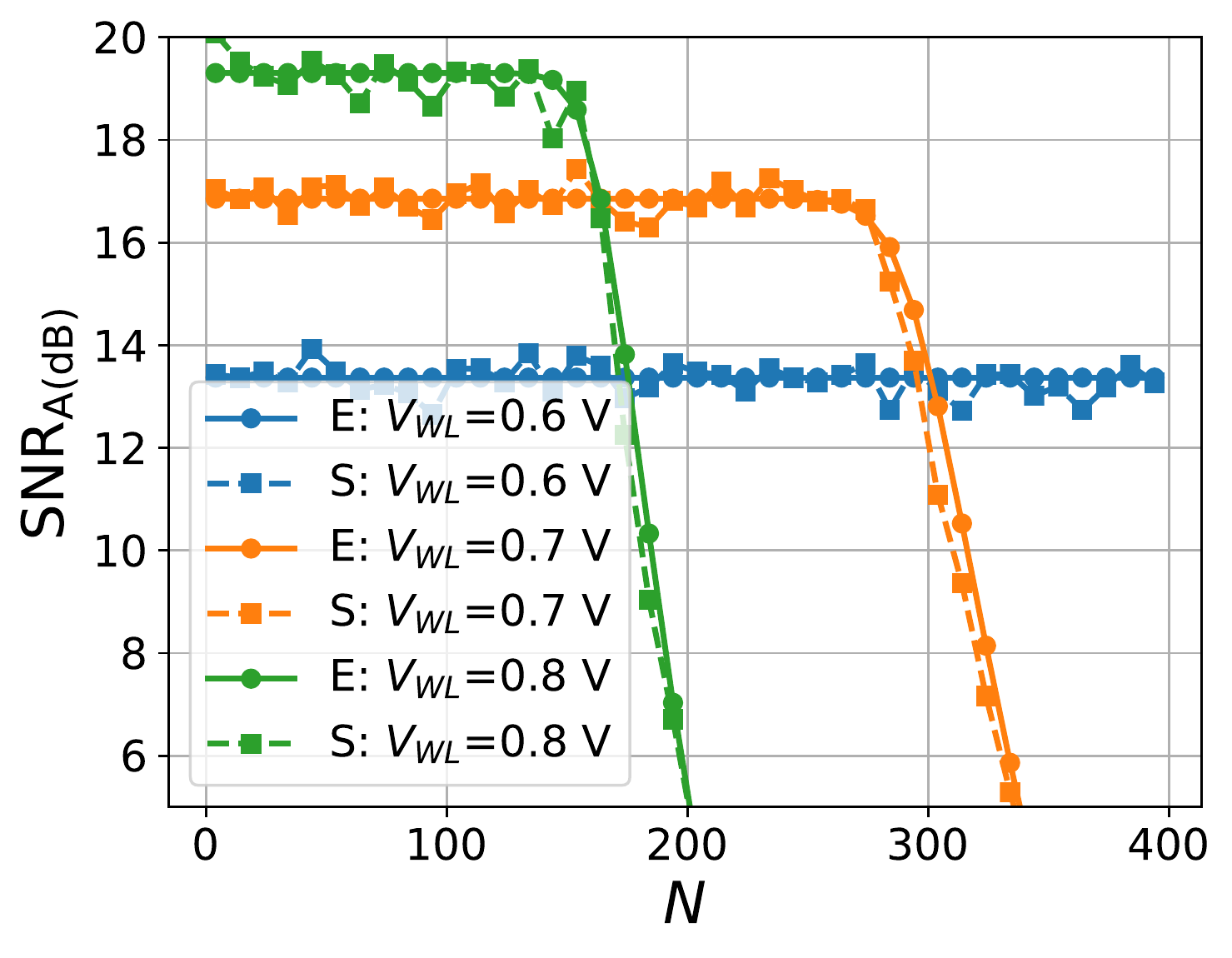}
        }\\
        \subfloat[]{
        \includegraphics[width=5cm]{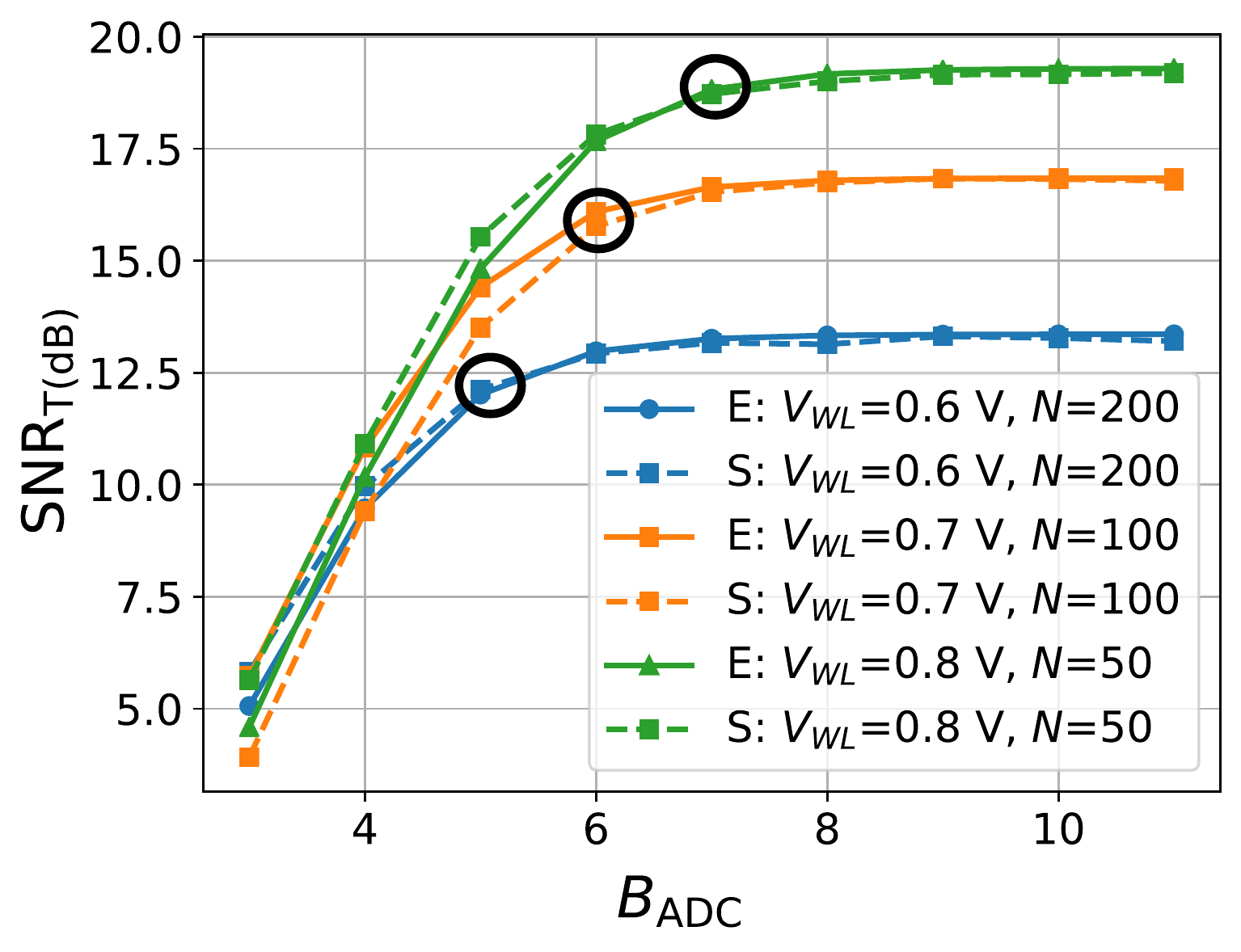}
        }
        \caption{$\text{SNR}$ trade-offs in the \QSArch{} with $B_x=B_w=6$: (a) $\SNRAdb$ vs. $N$ for different values of $V_{\text{WL}}$, and (b) $\SNRTdb$ vs. $B_\text{ADC}$  showing that the expression in Table \ref{tab:arch_summary} correctly predicts the minimum ADC precision $B_\text{ADC}$ (circled). Close match is achieved between expressions in Table~\ref{tab:arch_summary} (E) and simulations (S) of (\ref{eqn:QS-noise-model}).}
        \label{fig:BSBPCHSU_SNR1}
    \end{minipage}\hfill
    \begin{minipage}{0.37\textwidth}\vspace{0pt}
        \centering
        \subfloat[]{
        \includegraphics[width=5cm]{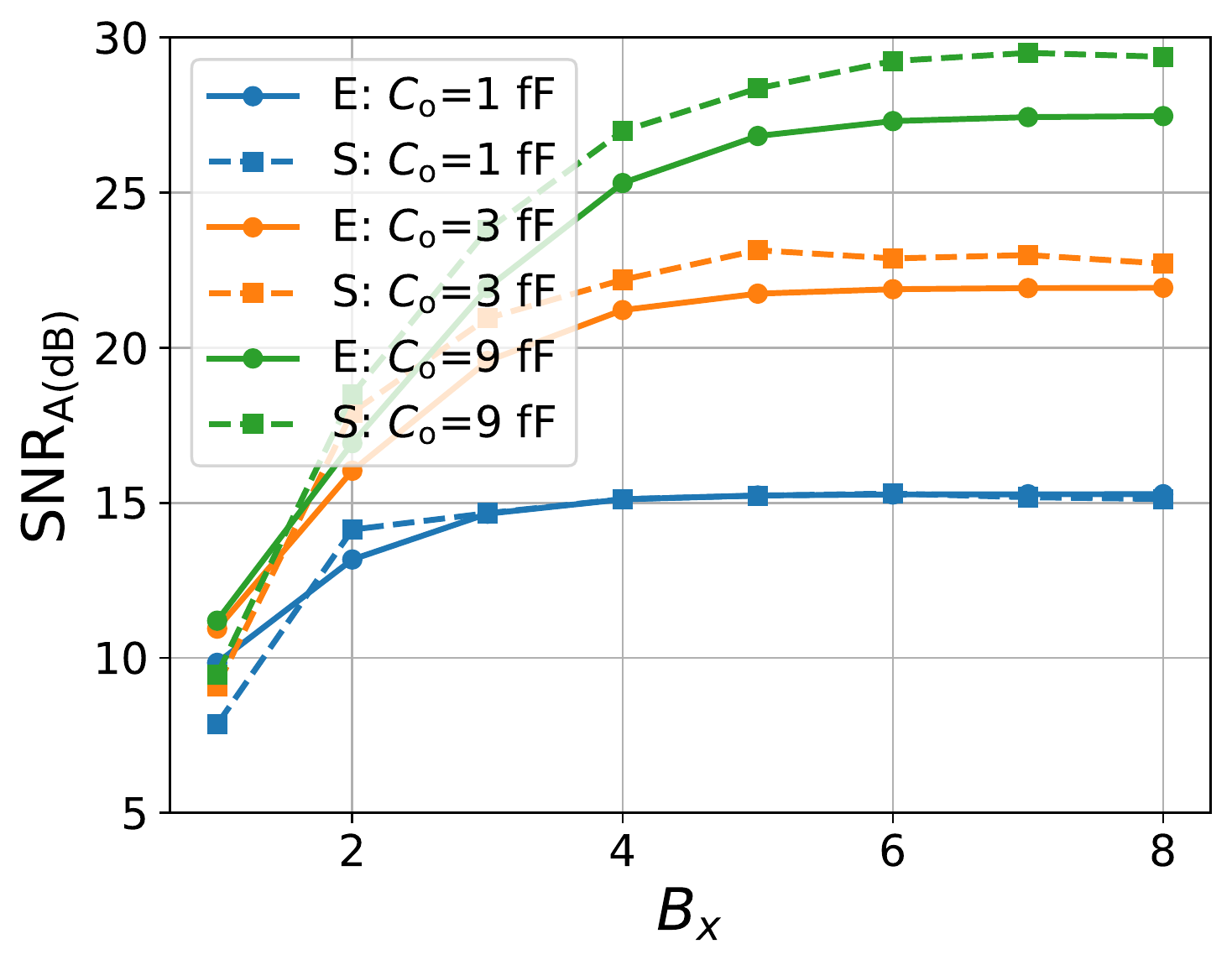}
        }\\
        \subfloat[]{
        \includegraphics[width=5cm]{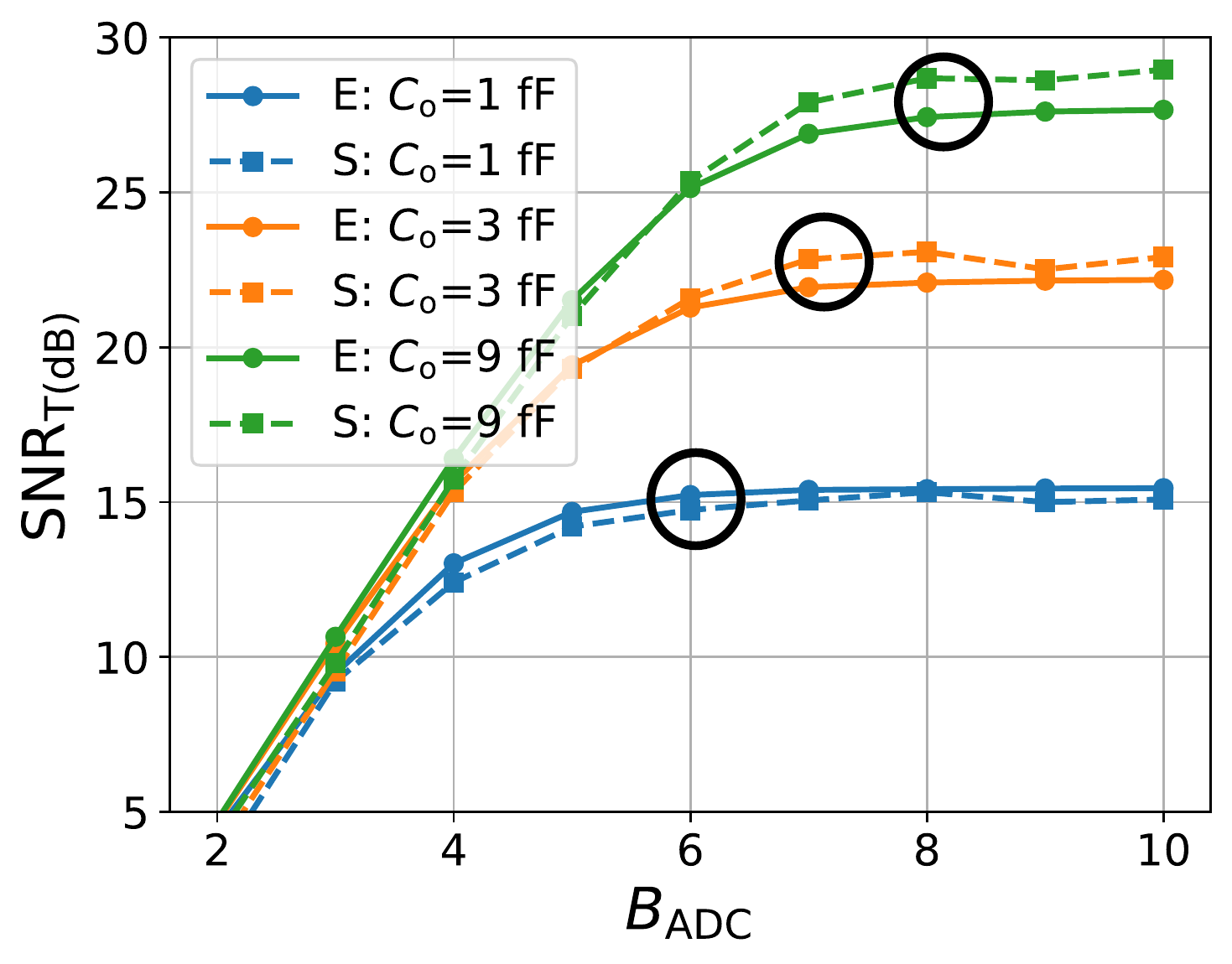}
        }
        \caption{$\text{SNR}$ trade-offs in the \QRArch{}    
     with $B_w=7$, and $N=64$: (a) $\SNRAdb$ as a function of $B_x$ for different values of $C_{\text{o}}$  showing that the SNR improves with $C_{\text{o}}$, and (b) $\SNRTdb$ as a function of $B_\text{ADC}$ for different values of $C_{\text{o}}$ with $B_x=6$ and $B_w=7$, showing that the expression in Table \ref{tab:arch_summary} correctly predicts the minimum ADC precision $B_\text{ADC}$ (circled). Here, `E' and `S' correspond to the evaluation of the expressions in Table~\ref{tab:arch_summary} and  sample-accurate simulations of (\ref{eqn:QR-noise-model}), respectively.}
        \label{fig:BM-ChSh-SNR}
    \end{minipage}\hfill
    \begin{minipage}{0.3\textwidth}\vspace{0pt}
        \centering
        \subfloat[]{
        \includegraphics[width=5cm]{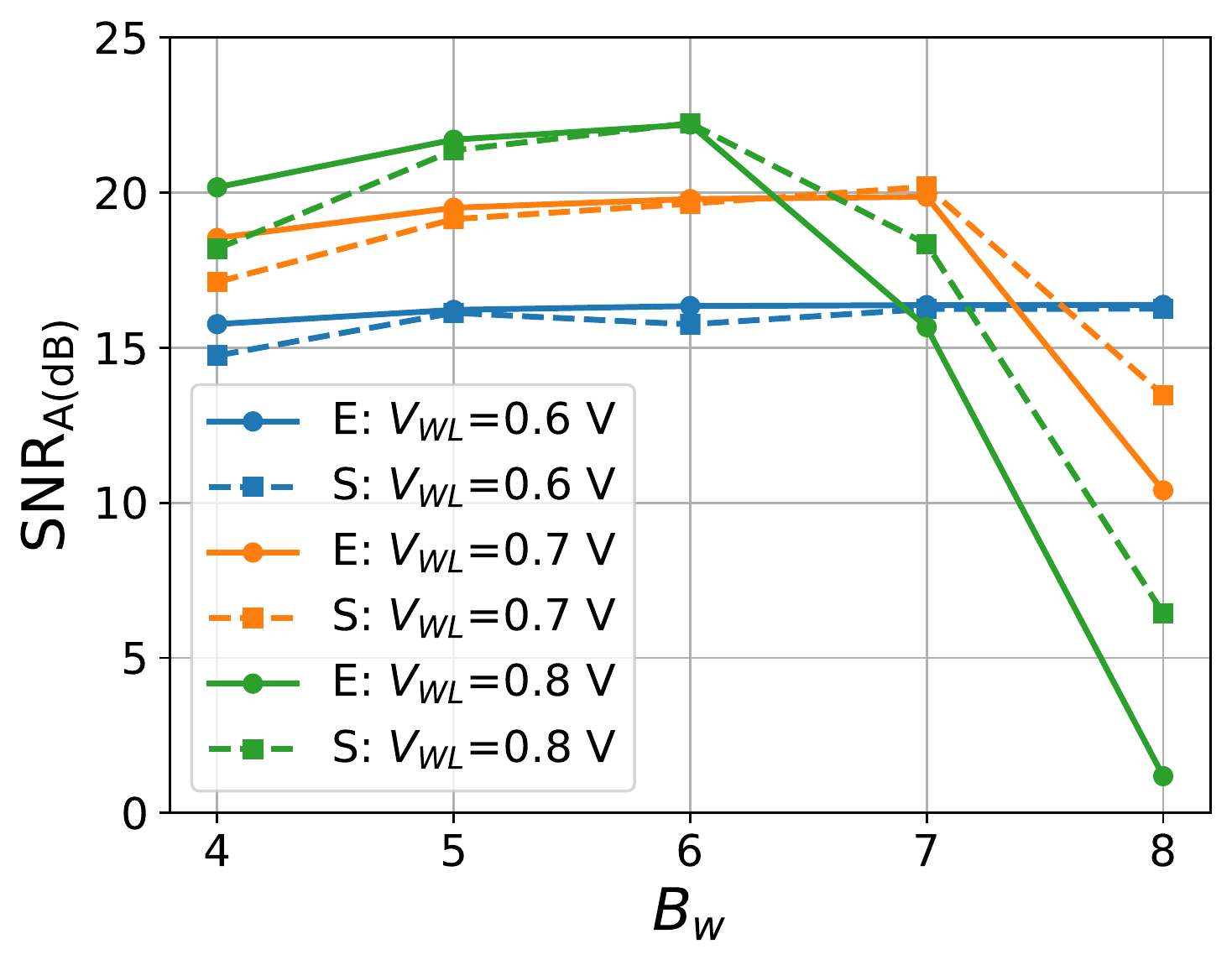}
        }\\
        \subfloat[]{
        \includegraphics[width=5cm]{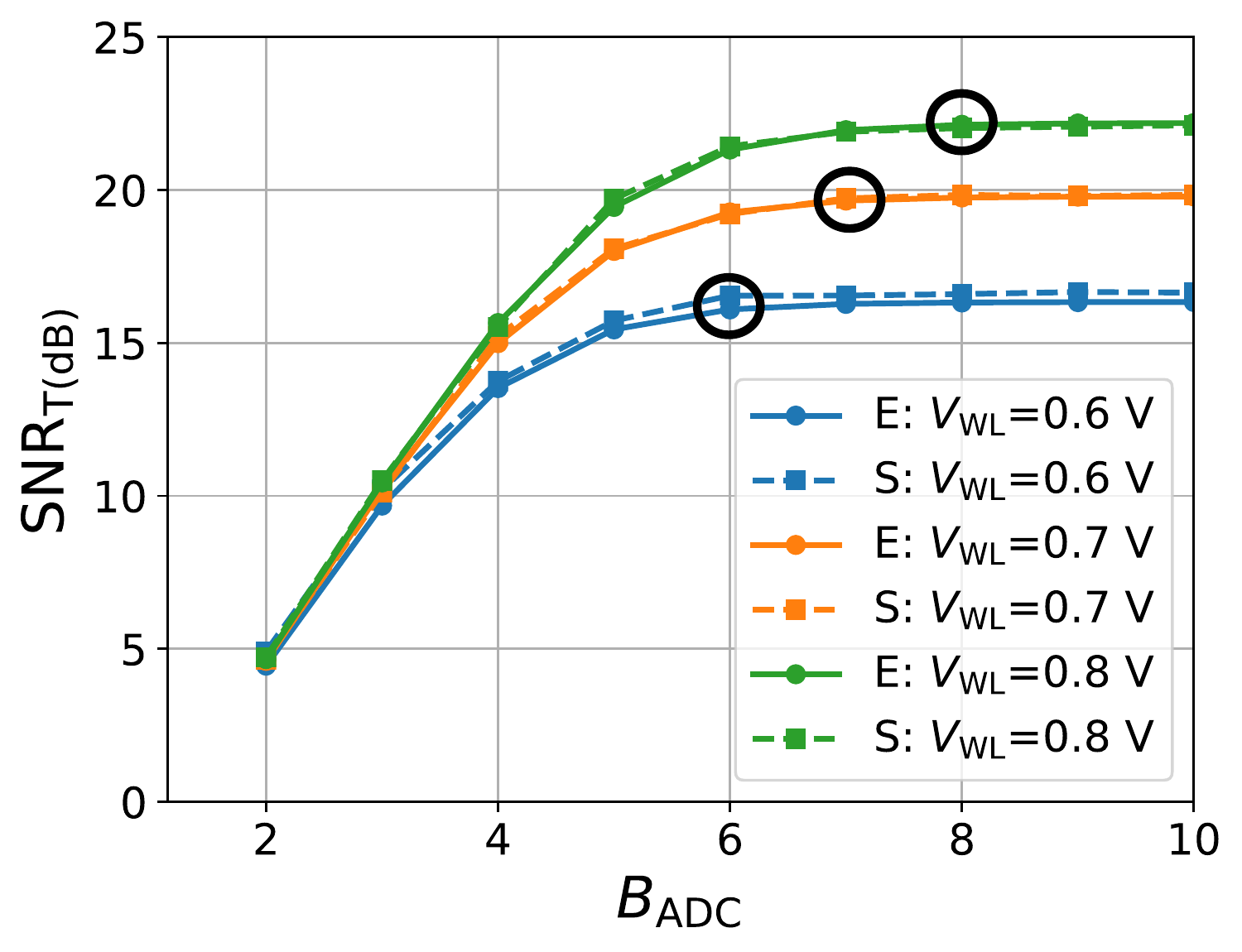}
        }
        \caption{SNR trade-offs in CM with $B_x=6$ and $N=128$: (a) $\SNRAdb$  vs. $B_w$  indicating the existence of an optimal value of $B_w$ that balances quantization and headroom clipping noise, and (b) $\SNRTdb$vs. $B_\text{ADC}$ with $B_w=6$. Here, `E' and `S' correspond to the evaluation of the expressions in Table~\ref{tab:arch_summary} and  sample-accurate simulations using (\ref{eqn:QS-noise-model}) and (\ref{eqn:QR-noise-model}), respectively.}
        \label{fig:CM_SNR}
    \end{minipage}\hfill
\end{figure*}
\subsubsection{ \QSArch{} }

Figure~\ref{fig:BSBPCHSU_SNR1}(a) shows that the maximum achievable $\SNRA$ increases with $V_\text{WL}$. Further, for a fixed $V_\text{WL}$, \QSArch{} also exhibits a sharp drop in $\SNRA$ at high values of $N>N_\text{max}$, e.g., $\SNRA\approx\unit[19.6]{dB}$ for $N\leq 125$ and then drops with increase in $N$. A key reason for this trade-off is that $\sigma^2_{\eta_{\text{h}}}$ decreases while $\sigma^2_{\eta_\text{e}}$ increases as $V_\text{WL}$ is reduced (see Table~\ref{tab:arch_summary}), and since $\sigma^2_{\eta_\text{h}}$ limits $N$ and $\sigma^2_{\eta_\text{e}}$ limits $\SNRA$.  Thus, by controlling $V_{\text{WL}}$, we can trade-off $N_\text{max}$ with $\SNRA$. Specifically, $N_\text{max}$ increases by $2\times$ for every \unit[3]{dB} drop in $\SNRA$.

%In \QSArch{} , $k_{\text{h}}$ and $\sigma_\text{D}$ both increase with $V_\text{WL}$, we observe a trade-off between $\sigma^2_{\eta_\text{h}}$ and $\sigma^2_{\eta_\text{e}}$ from their expressions in Table~\ref{tab:arch_summary}, i.e., $\sigma^2_{\eta_{\text{h}}}$ decreases while $\sigma^2_{\eta_\text{e}}$ increases as $V_\text{WL}$ is reduced. Since $\sigma^2_{\eta_\text{h}}$ limits $N$ and $\sigma^2_{\eta_\text{e}}$ limits $\text{SNR}_{\text{a}}$, \QSArch{} exhibits a trade-off between the DP dimension $N$ and $\text{SNR}_{\text{a}}$ as seen in Fig.~\ref{fig:BSBPCHSU_SNR1}(a).

In \QSArch{}, the minimum value of $B_\text{ADC}$ (see Table~\ref{tab:arch_summary}) depends upon the minimum of: 1) the MPC term (\ref{eqn:bympc}); 2) the headroom clipping term; and 3) the small $N$ case where BL discharge $\Delta V_{\text{BL}}$ has a finite number of discrete levels.  
Figure~\ref{fig:BSBPCHSU_SNR1}(b) shows that $\SNRT\rightarrow\SNRA$ of Fig.~\ref{fig:BSBPCHSU_SNR1}(a) when $B_\text{ADC}$ is greater than the lower bound (circled) in Table~\ref{tab:arch_summary} for different values of $V_{\text{WL}}$ and $N$.

\color{black}
\subsubsection{ \QRArch{} }

\QRArch{} demonstrates a clear energy-accuracy-area trade-off as seen in Fig.~\ref{fig:BM-ChSh-SNR}(a). Here, $\SNRA$ improves with capacitor $C_{\text{o}}$ size but at the expense of higher energy and area costs. For instance, increasing $C_{\text{o}}$ from \unit[1]{fF} to \unit[3]{fF} and \unit[9]{fF} leads to $\text{SNR}_{\text{a}}$ improvements of $\sim\unit[8]{dB}$ and $\sim\unit[12]{dB}$, respectively. 

Figure~\ref{fig:BM-ChSh-SNR}(b) shows that the expressions in Table~\ref{tab:arch_summary} correctly predict the minimum value of ADC precision $B_\text{ADC}$ and the input range $V_{\text{c}}$. MPC is demonstrated to greatly reduce the ADC precision requirements as for this example 6-8 bits suffice in order to maintain $\text{SNR}_{\text{a}}$. In contrast, if BGC were to be employed, $B_\text{ADC}=12$ would have been assigned.

\subsubsection{ Compute Memory}

Figure~\ref{fig:CM_SNR}(a) shows that the quantization noise term $\sigma^2_{q_{iy}}$ reduces and the headroom clipping noise $\sigma^2_{\eta_{\text{h}}}$ increases as a function of $B_w$ implying an $\text{SNR}_{\text{a}}$-optimal value for 
$B_w$, e.g., $\text{SNR}_{\text{a}}$ peaks at $B_w=6$ and $B_w=7$ for $V_{\text{WL}}=\unit[0.8]{V}$ and $V_{\text{WL}}=\unit[0.7]{V}$, respectively.

Figure~\ref{fig:CM_SNR}(a) also shows another interesting trade-off, this time between headroom clipping noise $\sigma^2_{\eta_{\text{h}}}$ and $\sigma^2_{\eta_{\text{e}}}$, e.g., when $B_w=7$, $\text{SNR}_{\text{a}}$ is dominated by $\eta_{\text{e}}$ when $V_{\text{WL}}=\unit[0.6]{V}$ and by $\eta_{\text{h}}$ when $V_{\text{WL}}=\unit[0.8]{V}$. Furthermore, both noise sources are balanced when $V_{\text{WL}}=\unit[0.7]{V}$. In fact, one can show that the clipping threshold $k_{\text{h}}$ is proportional to $\sigma_\text{D}$ indicating this relationship.

 Figure~\ref{fig:CM_SNR}(b) shows that choosing $B_\text{ADC}$ using MPC (\ref{eqn:bympc}) ensures that the $\text{SNR}_{\text{T}}$ is indeed within \unit[0.5]{dB} of $\text{SNR}_{\text{a}}$ in Fig.~\ref{fig:CM_SNR}(a). Once more, MPC assigns $B_\text{ADC}\leq 8$ when $B_x=B_w=6$ and $N=128$ which is much smaller than $B_\text{ADC}=19$ determined via BGC.
 
\color{black}

\subsection{Impact of ADC Precision}
\begin{figure*}[!t]
    \centering
    \subfloat[]{
    \includegraphics[width=0.3\linewidth]{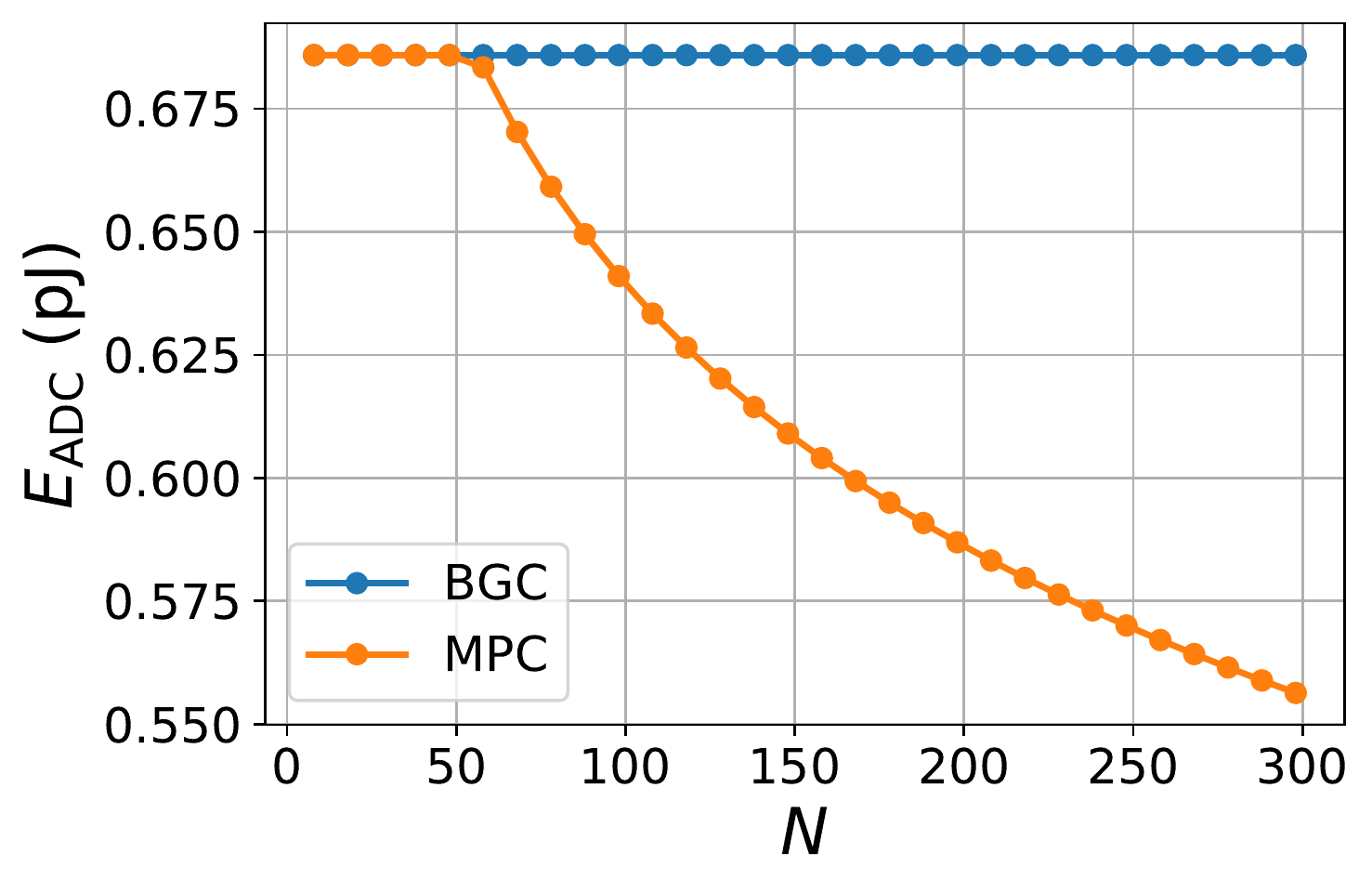}
    }
    \subfloat[]{
    \includegraphics[width=0.3\linewidth]{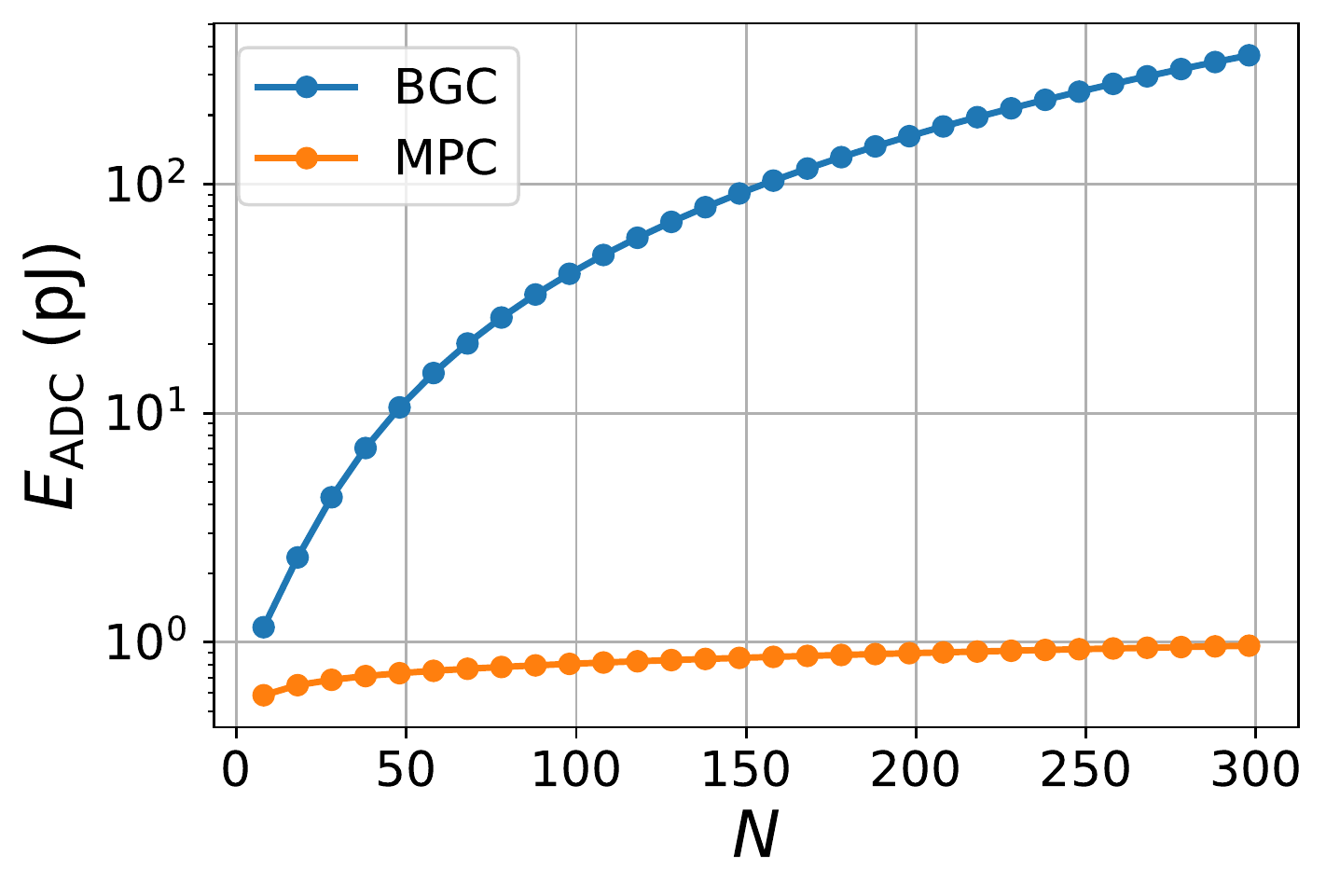}
    }
    \subfloat[]{
    \includegraphics[width=0.3\linewidth]{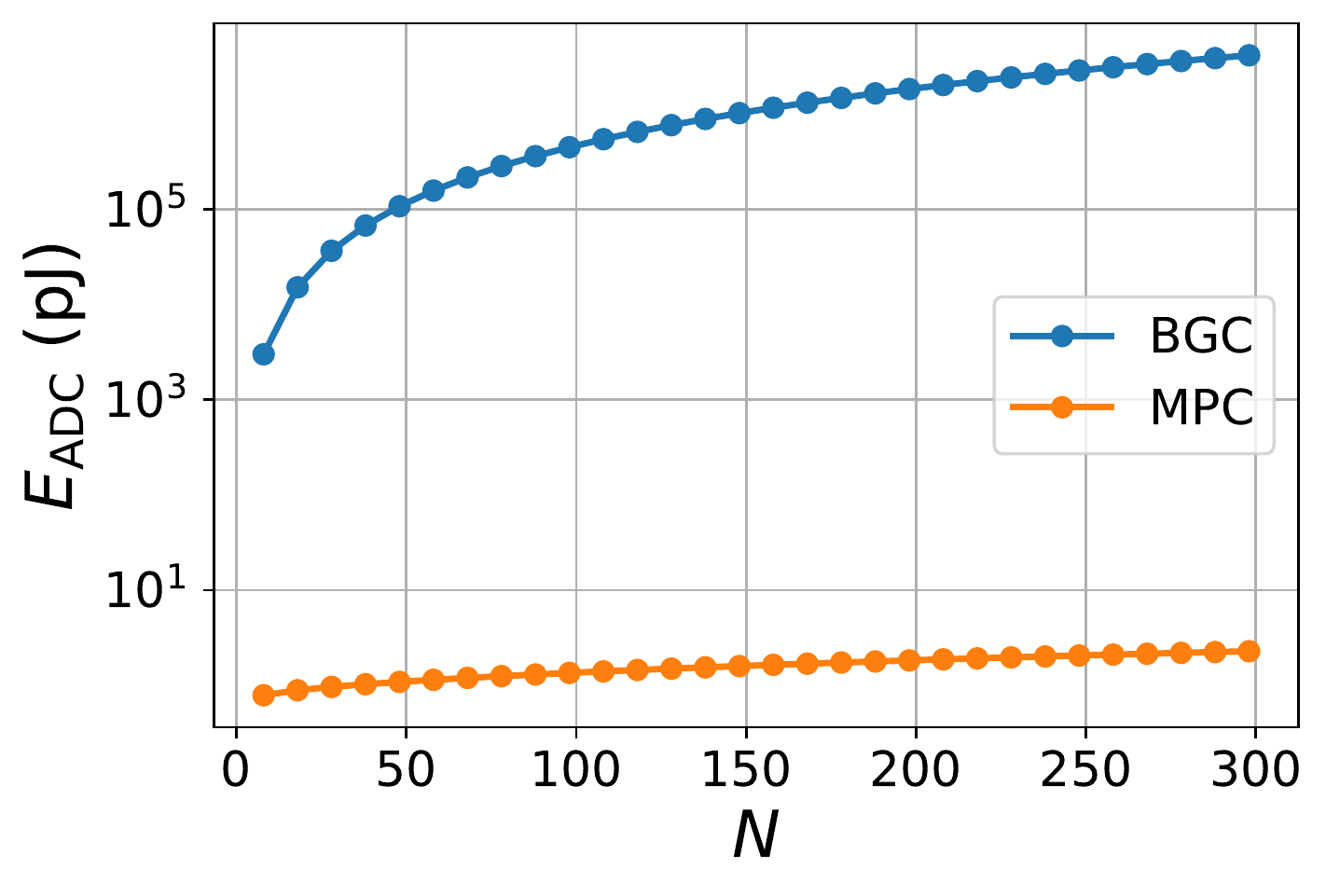}
    }
    \caption{ADC energy in (a) \QSArch{}, (b) \QRArch{} and (c) CM with $B_x=B_w=6$  as a function of $N$ when $B_\text{ADC}$ chosen according to BGC \eqref{eqn:bgc_assignment}, and the MPC criterion (Table~\ref{tab:arch_summary}) such that $\SNRTdb$ is within $\unit[0.5]{dB}$ of $\SNRAdb$. $V_{\text{WL}}=\unit[0.8]{V}$ in CM, $V_{\text{WL}}=\unit[0.7]{V}$ in \QSArch{}, and $C_{\text{o}}=\unit[3]{fF}$ in \QRArch{}.}
    \label{fig:ADC}
\end{figure*}

\begin{comment}
\color{blue}
\begin{figure}[!t]
    \centering
    \includegraphics[width=0.9\linewidth]{Figures/BSBPCHSU_plots/ADC_energy.pdf}
    
    \caption{ADC energy in \QSArch{} with $B_x=B_w=6$ at $\SNRAdb=\unit[16.2]{dB}$ as a function of $N$ when $B_\text{ADC}$ chosen according to BGC \eqref{eqn:bgc_assignment}, tBGC, and the MPC criterion (Table~\ref{tab:arch_summary}) such that $\SNRTdb$ is within $\unit[0.5]{dB}$ of $\SNRAdb$.}
    \label{fig:ADC}
\end{figure}

ADC energy costs when operating in the noise-limited regime is modeled as \cite{murmann2008d,murmann2015race}:
\begin{align}
    E_\text{ADC}=\beta 4^{B_\text{ADC}}
    \label{eqn:adc-energy}
\end{align}
where  $\beta$ is estimated from the Schreier figure of merit \cite{schreier2005understanding,murmann2015race} which is approximately \unit[180]{dB} based on recent (2019) ADCs \cite{ADCSurvey} leading to $\beta=\unit[7.5\times 10^{-4}]{fJ}$ at $V_{\text{dd}}=\unit[1]{V}$. 

Figure~\ref{fig:ADC} shows that ADC energy increases with DP dimension $N$. However, the gap between ADC energy consumption with BGC and MPC begins to increase for $N>60$. This is because BGC assigns higher values of $B_\text{ADC}$ as compared to MPC (see Table~\ref{tab:arch_summary}) to achieve the same $\SNRT$.

\end{comment}

Minimizing the column ADC energy is critical to maintain IMC's energy efficiency. ADCs in IMCs need to operate in a noise-limited regime due to the high PAR of high-dimensional DP outputs combined with severe area constraints imposed by column-pitch matching requirements. To estimate the ADC energy costs we use the following empirical model based on \cite{murmann2020tvlsi}:
\begin{align}
    E_\text{ADC}=k_1 \Big(B_{\text{ADC}}+\log_2\Bigg( \frac{V_\text{DD}}{V_{\text{c}}}\Big) \bigg) +k_2 \Big( \frac{V_{\text{DD}}}{V_{\text{c}}}\Big)^2 4^{B_{\text{ADC}}}
    \label{eqn:ADCenergy}
\end{align}
where $V_{\text{c}}$ is the voltage range that need to be quantized, and  $k_1=\unit[100]{fJ}$ and $k_2=\unit[1]{aJ}$ are empirical parameters \cite{murmann2020tvlsi} based on the recent ADCs \cite{ADCSurvey,murmann2008d}. 

From (\ref{eqn:ADCenergy}), it is clear that the energy consumption of ADC decreases with $V_{\text{c}}$ and increases with $B_{\text{ADC}}$. If BGC is employed, then $2^{B_{\text{ADC}}} \propto N$ (\ref{eqn:bgc_assignment}) resulting in ADC energy increasing with $N$ when $V_{\text{c}}$ is constant, as in the case of \QRArch{} and in CM (see Fig~\ref{fig:ADC}(b) and Fig~\ref{fig:ADC}(c)). However, in \QSArch{}, $V_\text{c} \propto N$ therefore the ADC energy consumption in \QSArch{} remains constant with $N$ (see Fig~\ref{fig:ADC}(a)).

On the other hand, if MPC is employed, $B_{\text{ADC}}$ remains constant with $N$ (Table~\ref{tab:arch_summary}), and hence $E_\text{ADC}$ only depends on $V_{\text{c}}$. In \QSArch{}, $E_\text{ADC}$ reduces with $N$ as $V_\text{c} \propto \sqrt{N}$, while in \QRArch{} and CM $E_\text{ADC}$ increases with $N$ as $V_\text{c} \propto 1/\sqrt{N}$ (see Fig~\ref{fig:ADC}(a)). Note that for \QRArch{} and CM, MPC leads to significant energy savings over BGC criterion since $E_\text{ADC} \propto N^2$ in BGC while  $E_\text{ADC} \propto N$ using MPC.
\begin{comment}
\begin{table}[]
    \centering
    \caption{Dot product energy consumption for $N=100$, $B_w=4$, and $B_x=3$ at $\SNRTdb = \unit[17]{dB}$ \sujan{Need to define IMC energy}}
    \begin{tabular}{|c|c|c|}
        \hline
            & IMC energy (pJ) & ADC energy (pJ) \\
        \hline
         \QSArch{}  &  2.2 & 7.2 \\
         \hline
         \QRArch{} & 0.95 & 3.1 \\
         \hline
          CM & 6.8 & 1.1 \\
         \hline
    \end{tabular}
    \label{tab:dot-energy}
\end{table}

\color{red}

Note that, though ADC energy in \QSArch{} reduces with $N$ it increased with $B_w$ and $B_x$ in \QSArch{} as it decomposes the dot-product into binarized dot products. Similarity, ADC energy in \QRArch{} increases with $B_w$.
This can be seen in Table~\ref{tab:dot-energy} where ADC energy in \QSArch{} is the highest followed by \QRArch{} and CM.
\end{comment}
%% SUJAN : I had not multiplied the ADC energy with the number of ADC operations here

%In fact, when $N=100$ and $\SNRTdb = \unit[17]{dB}$, the ADC energy is approximately $20\%$, $40\%$, and $16\%$ of the total energy in \QSArch{}, \QRArch{}, and CM, respectively.

\color{black}

%The cost of ADC becomes particularly important while operating with a small $V_{\text{c}}$, since high-precision ADCs are required. ADCs that are pitch matched to the array are typically noise-limited as they operate with very limited area budget. 

%For example, an ADC with \unit[1]{mV} quantization step will require the thermal noise floor to be in the order of \unit[0.3]{mV} to guarantee accurate conversion with $>$99\% probability, thus requiring capacitors in the order of \unit[100]{fF}. Such a capacitor needs the same area as 20 6T SRAM bit-cells. Energy costs of noise-limited ADC designs can be modeled \cite{murmann2010trends,murmann2008d} as:

%where $\Delta_\text{ADC}$ is the ADC step size in voltage domain, $V_{\text{c}}$ is the voltage range that need to be quantized, and $\beta$ can be estimated from the Schreier figure of merit \cite{schreier2005understanding,murmann2015race}. Considering the recent ADCs tabulated in \cite{ADCSurvey}, the best Schreier figure of merit is about \unit[180]{dB} as of the year 2019, leading to $\beta=\unit[7.5\times 10^{-4}]{fJV^2}$ with $V_{\text{dd}}=\unit[1]{V}$. 
%\begin{align}
%    E_\text{ADC}=\frac{\beta}{\Delta_\text{ADC}^2}=\frac{\beta}{V_{\text{c}}^2}4^{B_y}
%\end{align}
%The ADC energy can easily dominate while operating with large DP lengths in architectures that use QR compute models. Table \ref{tab:arch_summary} shows that $V_{\text{c}}$ reduces with increasing $N$ in CM and \QRArch{}, respectively.

\subsection{Impact of Technology Scaling}
\begin{figure*}[!t]
    \centering    
    \subfloat[]{
    \includegraphics[width=0.3\linewidth]{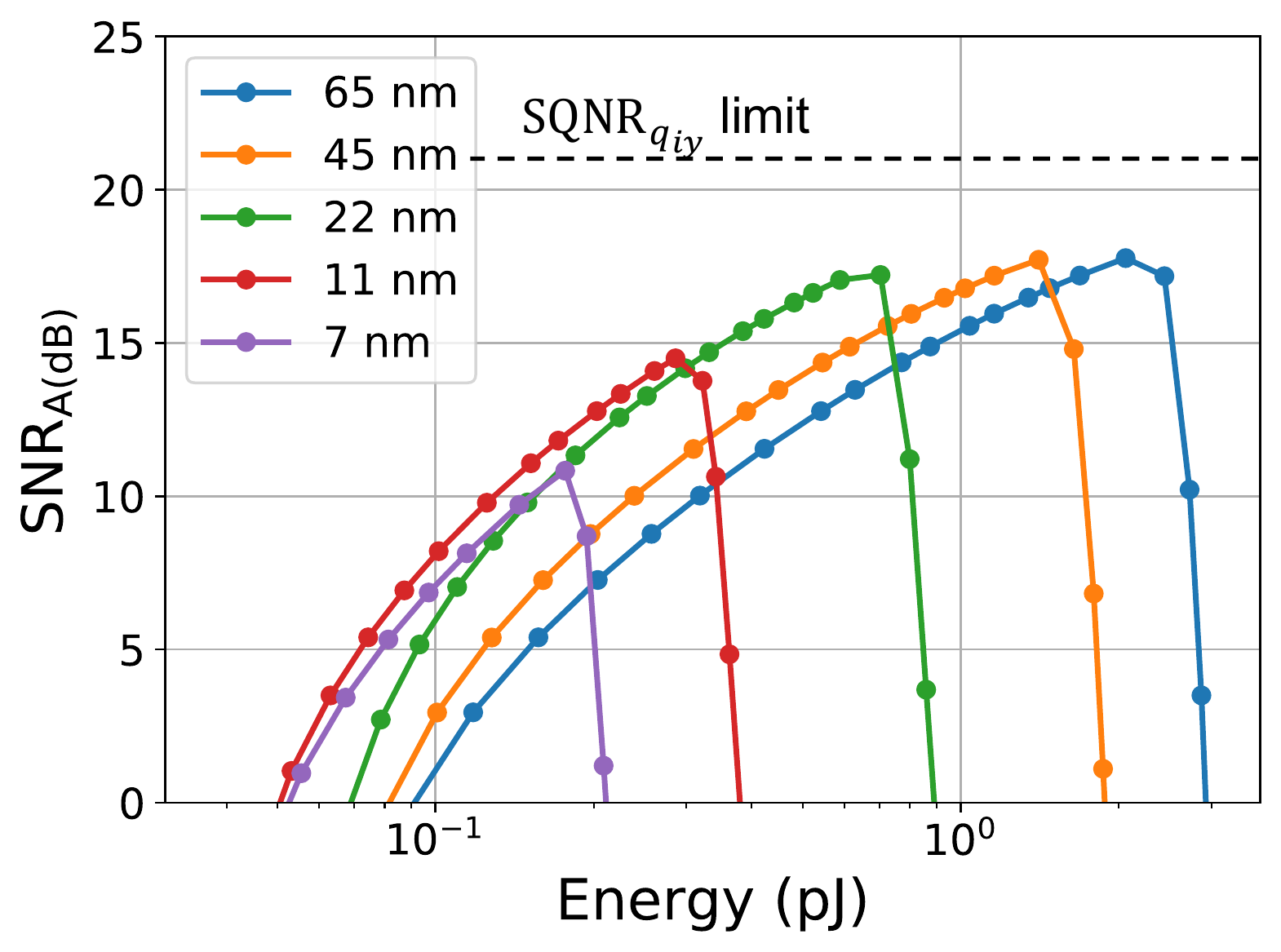}
    }
    \subfloat[]{
    \includegraphics[width=0.3\linewidth]{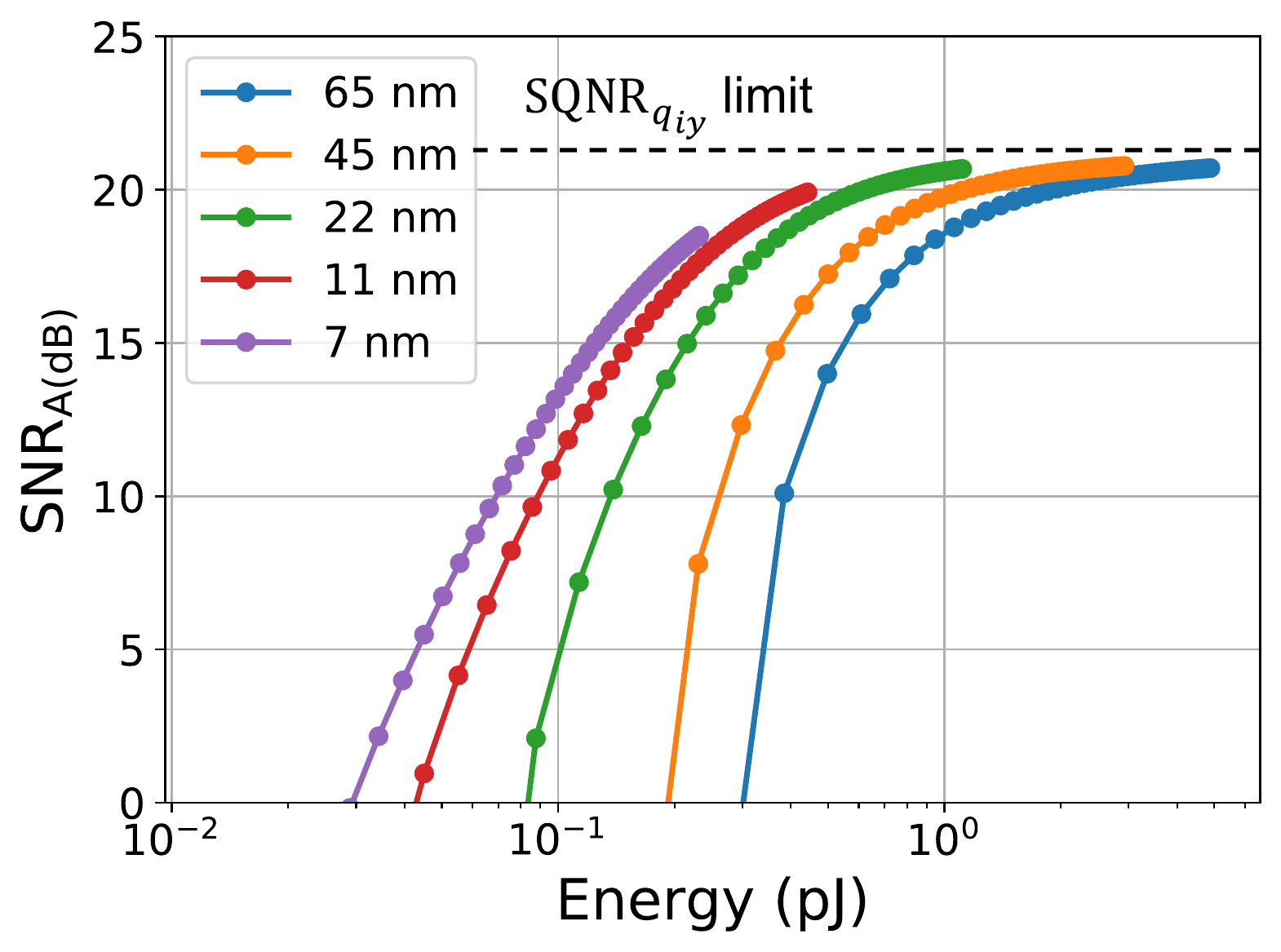}
    }
    \subfloat[]{
    \includegraphics[width=0.3\linewidth]{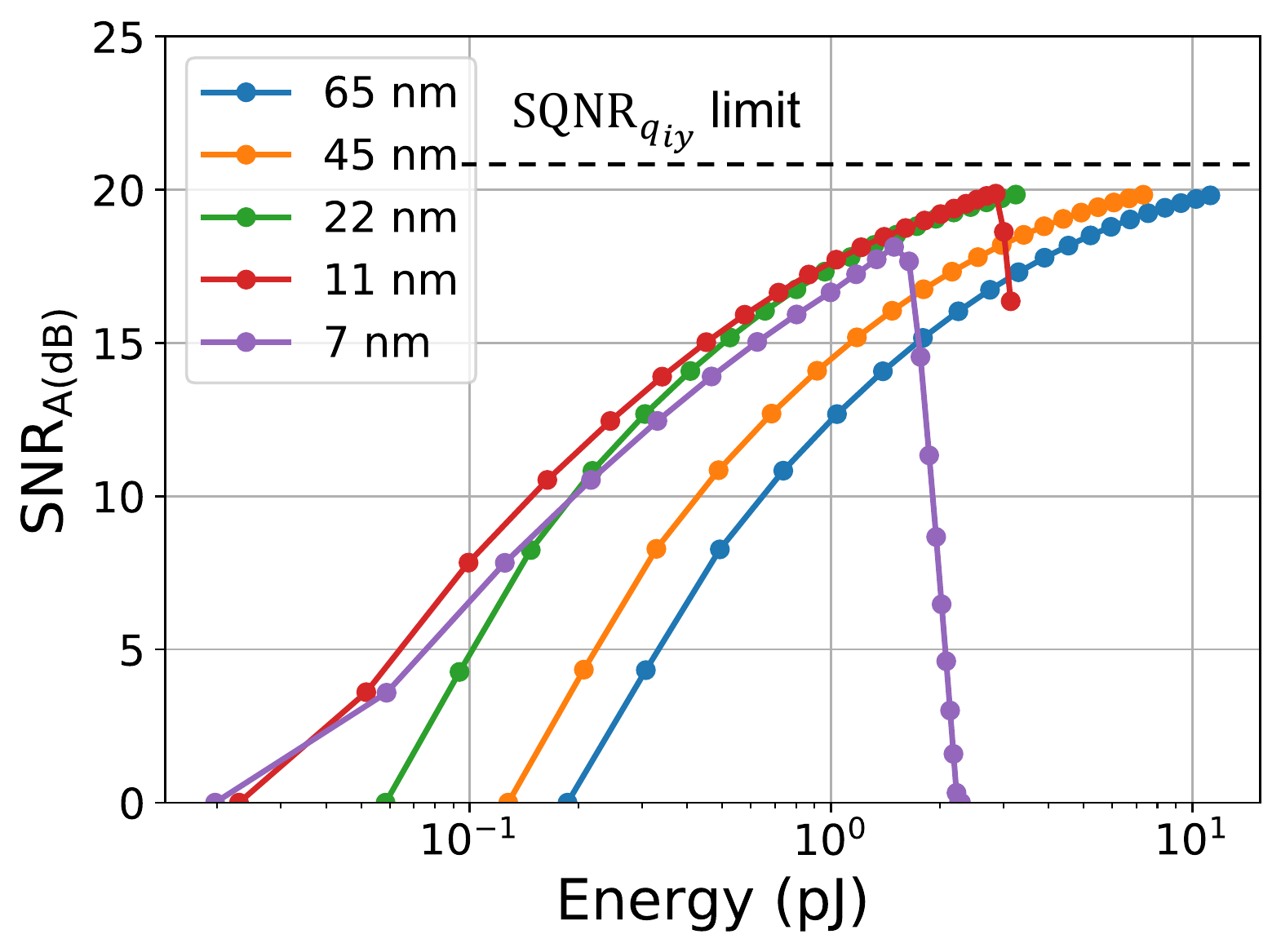}
    }
    \caption{Impact of CMOS technology scaling on the compute SNR vs. energy trade-off in  (a) \QSArch{} (swept parameter: $V_\text{WL}$), (b) \QRArch{} (swept parameter $C_{\text{o}}$), and  (c) CM (swept parameter: $V_\text{WL}$) with $B_x=3$, $B_w=4$, and $N=100$.
    }
    \label{fig:TechTrend}
\end{figure*}
One expects IMCs to exhibit improved energy efficiency and throughput in advanced process nodes due to lower capacitance and lower supply voltage. However, the impact of technology scaling on the analog noise sources also needs to be considered. To study this trade-off, we employ the SNR and energy models from Section~\ref{sec:compute-models} (see Table \ref{tab:arch_summary}) with parameters scaled as per the ITRS roadmap \cite{ITRS}. FDSOI technology is assumed for the $\unit[22]{nm}$, $\unit[11]{nm}$ and $\unit[7]{nm}$ nodes.

For a specific node, Fig.~\ref{fig:TechTrend} shows that the  energy cost reduces by $2\times$ in CM and \QSArch{}, and  $4\times$ in \QRArch{} for every $\unit[6]{dB}$ drop in $\SNRA$. \QSArch{} suffers a catastrophic drop in $\SNRA$ before reaching the input quantization noise limit set by \eqref{eqn:sqnrqiy}. This drop occurs due to an increase in the clipping noise variance $\sigma^2_{\eta_\text{h}}$. \color{black} In contrast, \QRArch{} is able to approach quantization noise limits as it does not suffer from headroom clipping noise.
\color{black}

Across technology nodes, the
maximum achievable $\SNRA$ in
\QSArch{} and CM \emph{reduces} as technology scales from $\unit[65]{nm}$ down to $\unit[7]{nm}$ due to: 1) increased clipping probability caused by lower supply voltages, and 2) increased variations in BL discharge voltage $\Delta V_{\text{BL}}$ due to smaller $V_{\text{dd}}/V_{\text{t}}$ ratio. As a result,
Fig.~\ref{fig:TechTrend} also shows that the energy consumption, \emph{at the same} $\SNRA$, is in fact \emph{higher} in \unit[11]{nm} and \unit[7]{nm} nodes as compared to the \unit[22]{nm} node in \QSArch{ and CM }due to the need to employ a higher values of $V_{\text{WL}}$ to control variations in $\Delta V_{\text{BL}}$ implying the technology scaling may not be friendly to IMCs based on the QS compute model.  %However, due to the increased wire-load capacitance and resistance, the throughput improvements will reduce for \unit[11]{nm} and \unit[7]{nm} nodes. 

%The reason for this is the increased WL resistance due to width-dependent scattering \cite{WLeffects} in $\leq \unit[16]{nm}$ technology nodes which limits the smallest WL pulse-width $T_0$ and thereby increases the headroom clipping probability and energy consumption. 

\section{Conclusions and Summary}
\label{sec:guidelines}
%Discussion in Section \ref{sec:arch-noise-models} and \ref{sec:energy} shows the relationships between the DP length $N$, the SNR, computation energy, and ADC energy. It is clear that the application requirements determine the choice of architectures and that there is not a single architectural choice that is optimal for all scenarios. Below are a few considerations for designing IMCs based on the applications requirements:
Based on the results presented in the earlier sections, we provide the following IMC design guidelines:%\vspace{-1em}
\begin{itemize}[leftmargin=*]
    \item For IMCs to be useful in realizing DNNs, the compute SNR of their analog core ($\SNRa$) needs to be the range ~$\unit[10]{dB}-\unit[40]{dB}$ or greater depending on the layer. This is because the total SNR ($\SNRT$) of DP computations implemented on IMCs is upper bounded by $\SNRa$. 
    \item In order for $\SNRT\rightarrow\SNRa$ with minimal energy and latency costs: 1) use \eqref{eqn:sqnrqiy} to ensure that the weight and activation precisions are sufficiently (e.g., $\unit[9]{dB}$) below analog noise sources; and 2) use the MPC-based Precision Assignment Rule \eqref{eqn:bympc} to assign the ADC precision.
    \item QS-based architectures tend exhibit lower energy cost at low compute SNRs. Meanwhile, QR-based architectures are preferred when the compute SNR requirements are higher.
    \item For the \QSArch{}, given an array size, there exists a trade-off between the maximum achievable $\SNRa$ and the maximum realizable DP dimension $N$. Multi-bank IMCs will be required for high-dimensional DPs in order to boost the overall compute SNR.
    \item Technology scaling will have an adverse impact on the maximum achievable $\SNRa$ and the energy cost incurred for a fixed $\SNRa$ for both \QSArch{} and CM. 
%    \item \QSArch{} allows the implementation of variable input and weight precision on the same hardware making it suitable for  exploiting energy-vs-accuracy trade-offs via compiler-driven methods such as those described in \cite{promise}.
   \item When MPC is employed, the ADC energy increases with the DP size $N$ for both \QRArch{} and CM due to the decreasing signal variance in the QR compute model. The opposite trend is observed for \QSArch{} where the ADC energy decreases as $N$ is increased.
   \item CM avoids incurring large ADC energy consumption by realizing multi-bit DPs instead of binarized ones.
   
   %  A rule of thumb in choosing the ADC step size is to set $\Delta_\text{ADC}=\sqrt{G/N}$, where $G$ is a constant that depends on the architecture, data statistics, and the SNR, e.g., it can be obtained using Table \ref{tab:arch_summary}.\nrs{apply to QSArch.}
\end{itemize}
An overarching conclusion of this paper is that the drive towards minimizing energy and latency using IMCs, runs counter to meeting the compute SNR requirements imposed by applications. This paper quantifies this trade-off through analytical expressions for compute SNR and energy-delay models. It is hoped that IMC designers will employ these models as they seek to optimize the design of IMCs of the future, including the use of algorithmic methods for SNR boosting such as statistical error compensation (SEC)  \cite{shanbhag2018shannon}.

\appendices
\section{SQNR expressions}
We present the derivation of the expressions of $\text{SQNR}_{q_{iy}(\text{dB})}$ in \eqref{eqn:sqnrqiy}, $\SQNRqydb$ in \eqref{eqn:sqnrqy}, $\SQNRqydb^{\text{BGC}}$ in \eqref{eqn:sqnr-bgc}, and $\SQNRqydb^{\text{MPC}}$ in \eqref{eqn:sqnr-mpc}.

\noindent\textbf{Derivation of $\text{SQNR}_{q_{iy}(\text{dB})}$ in \eqref{eqn:sqnrqiy}:}\\
Substituting $\Delta_w = w_\text{m}2^{-B_w+1}$ and $\Delta_x = x_\text{m}2^{-B_x}$ in \eqref{eqn:quant-noise} yields:
\begin{align}
    \label{eqn:sigma_qiy_detail}
    \sigma^2_{q_{iy}} = \frac{N}{3}\left(\sigma_w^2\frac{x_m^2}{4}2^{-2B_x}+\mathbb{E}[x^2]w_m^22^{-2B_w}\right)
\end{align}
which we substitute into the expression of $\SQNRqiy$ in \eqref{eqn:imc-accuracy-metrics} along with the expression of $\sigma_{y_{\text{o}}}^2$ from \eqref{eqn:quant-noise} to obtain:
\begin{align}
    \label{eqn:sqnrqiy_detail}
    \SQNRqiy = \frac{N\mathbb{E}[x^2]\sigma_w^2}{\frac{N}{3}\left(\sigma_w^2\frac{x_m^2}{4}2^{-2B_x}+\mathbb{E}[x^2]w_m^22^{-2B_w}\right)}
\end{align}
Dividing both numerators and denominators by $\frac{N}{3}\mathbb{E}[x^2]\sigma_w^22^{-2(B_x+B_w)}$, \eqref{eqn:sqnrqiy_detail} can be written as:
\begin{align}
    \label{eqn:sqnrqiy_final}
    \SQNRqiy = \frac{3\times2^{2(B_x+B_w)}}{\zeta_x^2\zeta_w^2\left(\frac{2^{2B_x}}{\zeta_x^2}+\frac{2^{2B_w}}{\zeta_w^2} \right)}
\end{align}
The result for $\text{SQNR}_{q_{iy}(\text{dB})}$ in \eqref{eqn:sqnrqiy} follows by taking $\text{SQNR}_{q_{iy}(\text{dB})} = 10\log_{10}(\text{SQNR}_{q_{iy}})$, with $\text{SQNR}_{q_{iy}}$ given by \eqref{eqn:sqnrqiy_final}.

\noindent\textbf{Derivation of $\SQNRqydb$ in \eqref{eqn:sqnrqy}:}\\
From the SQNR definition in \eqref{eqn:sqnr_db}, we have
\begin{align*}
    \SQNRqydb=6B_y +4.78 -\zeta_{y(\text{dB})}
\end{align*}
Thus, it suffices to show that $\zeta_{y(\text{dB})} = \zeta_{x(\text{dB})}+\zeta_{w(\text{dB})}+10\log_{10}(N)$ for the result in \eqref{eqn:sqnr_db} to follow. For simplicity, let us assume signed inputs and weights. Since $y_{\text{o}}=\mathbf{w}^{\mathsf{T}}\mathbf{x}$, we have
$y_m = Nx_mw_m$ (no clipping) and $\sigma_{y_{\text{o}}} = \sqrt{N}\sigma_x\sigma_w$.
Thus,
$\zeta_y = \frac{y_m}{\sigma_{y_{\text{o}}}}=\sqrt{N}\zeta_x\zeta_w$ with $\zeta_x = \frac{x_m}{\sigma_x}$ and $\zeta_w=\frac{w_m}{\sigma_w}$. The result follows by writing $\zeta_{y(\text{dB})}=20\log_{10}(\zeta_y)$.

\noindent\textbf{Derivation of $\SQNRqydb^{\text{BGC}}$ in \eqref{eqn:sqnr-bgc}:}\\
The result follows by replacing $B_y$ in \eqref{eqn:sqnrqy} by $ B_x + B_w + \log_2(N)$ which is assigned by BGC as per \eqref{eqn:bgc_assignment}. Note that $6\log_2(N) \approx 20 \log_{10}(N)$ hence we obtain
\begin{align*}
    \SQNRqydb^{\text{BGC}}
    &= 6(B_x+B_w+\log_2(N))  \\ &\quad\quad -  \zeta_{x(\text{dB})}-\zeta_{w(\text{dB})}-10\log_{10}(N)\\
    &\approx  6(B_x+B_w) + 20\log_{10}(N)\\ &\quad\quad - \zeta_{x(\text{dB})}-\zeta_{w(\text{dB})}-10\log_{10}(N)\\
    &= 6(B_x+B_w) - \zeta_{x(\text{dB})}-\zeta_{w(\text{dB})}+10\log_{10}(N)
\end{align*}
as listed in \eqref{eqn:sqnr-bgc}.

\noindent\textbf{Derivation of $\SQNRqydb^{\text{MPC}}$ in \eqref{eqn:sqnr-mpc}:}\\
In MPC we have:
\begin{align*}
    \SQNRqy^{\text{MPC}} = \frac{\sigma_{y_{\text{o}}}^2}{\sigma^2_{q_y}+p_c\sigma^2_{cc}} = \frac{\sigma_{y_{\text{o}}}^2}{\sigma^2_{q_y}\left(1+p_c\frac{\sigma^2_{cc}}{\sigma^2_{q_y}}\right)}
\end{align*}
with $\sigma^2_{q_y}=\frac{\Delta_y^2}{12}= \frac{y_c^22^{-2B_y}}{3}$, $p_{\text{c}}=\Pr\{|y_o|>y_\text{c}\}$, and $\sigma^2_{cc} =\mathbb{E}\big[\left(y_{\text{o}}-y_\text{c}\right)^2\big||y_{\text{o}}|>y_\text{c}\big]$. The above can be re-written as
\begin{align}
\label{eqn:sqnr_mpc_detail}
    \SQNRqy^{\text{MPC}} = \frac{3\times2^{2B_y}}{\left(\zeta_y^{\text{MPC}}\right)^2\left(1+p_c\frac{\sigma^2_{cc}}{\sigma^2_{q_y}}\right)}
\end{align}
with $\zeta_y^{\text{MPC}} = \frac{y_c}{\sigma_{y_{\text{o}}}^2}$. The result for $\SQNRqydb^{\text{MPC}}$ in \eqref{eqn:sqnr-mpc} follows by taking $\SQNRqydb^{\text{MPC}} = 10\log_{10}(\SQNRqy^{\text{MPC}})$, with $\SQNRqy^{\text{MPC}}$ given by \eqref{eqn:sqnr_mpc_detail}.

\section{Analog noise models expressions}
We present the derivation of expressions (\ref{eqn:current-mistmatch}), (\ref{eqn:rise-fall-times}), % (\ref{eqn:clipping_noise_variance}), (\ref{eqn:electrical_noise_cm}), (\ref{eqn:electrical_noise_bsbp}), and (\ref{eqn:VR-\QRArch{}}), 
expressions for noise variances ($\eta_{\text{h}}$, $\eta_{\text{e}}$) and ADC input range $V_{\text{c}}$ listed in Table \ref{tab:arch_summary}.

\noindent\textbf{Derivation of (\ref{eqn:current-mistmatch}):}\\
Employ the $\alpha$-law transistor I-V equation below to model the SRAM cell current $I_j$ (see Fig.~\ref{fig:cell_pulse_model}(a)):
\begin{align}
    I_{j} = \frac{W}{L}k'(V_{\text{WL}}-V_{\text{t}})^{\alpha}
    \label{eqn:alpha-law}
\end{align}
In the presence of threshold voltage variations, \eqref{eqn:alpha-law} transforms into:
\begin{align}
    I_{j}+i_j = \frac{W}{L}k'(V_{\text{WL}}-(V_{\text{t}}+v_{\text{t}}))^{\alpha}
    \label{eqn:alpha-law-variation}
\end{align}
where $v_{\text{t}}$ is the 
threshold voltage variation and $i_j$ is the resulting cell current variation. Using a $1^\text{st}$-order Taylor series expansion, we get:
\begin{align}
     i_j \approx v_\text{t}\frac{\partial I_j}{\partial V_\text{t}} %\nonumber\\
    %\frac{W}{L}k'(V_{\text{WL}}-V_{\text{t}}-\delta V_\text{t})^{\alpha} \nonumber\\
    %&= - \delta V_{\text{t}}\big(\alpha \frac{W}{L}k'(V_{\text{WL}}-V_{\text{t}})^{\alpha-1}\big) %\nonumber\\
    = - v_{\text{t}} \frac{\alpha I_j}{V_{\text{WL}}-V_{\text{t}}}
\end{align}
Assuming $v_{\text{t}}$ is a zero mean random variable with standard deviation $\sigma_{V_{t}}=\sqrt{\text{Var}(v_{\text{t}}))}$ leads to (\ref{eqn:current-mistmatch}) with
$\sigma_{I_{j}}=\sqrt{\text{Var}(i_j)}$.

\noindent\textbf{Derivation of (\ref{eqn:rise-fall-times}):}\\
To model the impact of finite rise/fall time on the total voltage discharge associated with $j$-th cell $ V_{\text{o},j}$, we integrate the SRAM cell current over the wordline pulse window $T_j$ to determine the total charge accumulated on the bitline cap $C$:
\begin{align}
 V_{\text{o},j} +v_j = \frac{1}{C} \int_{t=0}^{t=T_j}I_j(t)\text{d}t
\end{align}
where $V_{\text{o},j} = I_jT_j/C$ is the total discharge assuming an ideal $V_{\text{WL}}$ pulse ($T_{\text{r}}=T_\text{f}=0 \rightarrow I_j(t) = I_j$), and $v_j$ is the voltage drop that accounts for these effects. Modeling the SRAM cell current as an ideal current source with value set by (\ref{eqn:alpha-law}), we get:
\begin{align}
\label{eqn:tr_tf_deriv}
 V_{\text{o},j} +v_j = \frac{Wk'}{LC} \int_{t=0}^{t=T_j}\big(V_{\text{WL}}(t) - V_{\text{t}}\big)^\alpha \text{d}t
\end{align}
To simplify the analysis, we employ a linear approximation (red curve) to a realistic  $V_{\text{WL}}$ waveform in Fig.~\ref{fig:cell_pulse_model}(b)). Evaluating \eqref{eqn:tr_tf_deriv} with the linear approximation results in:
\begin{align}
 V_{\text{o},j} +v_j = \frac{I_{j}}{C}\Big[T_j-T_\text{r}+\Big(\frac{V_{\text{WL}}-V_\text{t}}{V_{\text{WL}}}\Big)\frac{T_\text{r}+T_\text{f}}{\alpha+1}\Big]
\end{align}
Thereby obtaining (\ref{eqn:rise-fall-times})
to account effect of rise and fall times of the WL pulse.

\noindent\textbf{Derivation of (\ref{eqn:qs-thermal})}
To derive the thermal noise variance in (\ref{eqn:qs-thermal}), we employ the bit-cell model in Fig~\ref{fig:cell_pulse_model}(a). The access transistors in the $j$-th bitcell contributes thermal noise $i_{\theta,j}$. The final BL thermal noise voltage $v_{\theta}$ is obtained by integrating the thermal noise current ($i_{\theta,j}$) contributions from the $N$ bit-cells attached to the BL, on the output capacitor $C$ as follows:
\begin{align}
    v_{\theta}=\frac{1}{C}\sum_{j=1}^N w_j \Bigg[ \int_{t=0}^{t=T_j} i_{\theta,j}\, \text{d}t \Bigg]
\end{align}

Assuming the access transistors are in saturation, the two-sided power spectral density of $i_{\theta,j}$ is given by $S_{i_{\theta}}(f)=\frac{4}{3} g_m k T$. Therefore:
\begin{align}
\mathbb{E}\Bigg[ \Bigg( \int_{t=0}^{t=T_j} i_{\theta,j}\, \text{d}t \Bigg)^2 \Bigg]= T_j S_{i_{\theta}}(0) = T_j \frac{4}{3} g_m k T 
\end{align}
We assume $T_j=x_jT_{\text{max}}$ are independent and the integrated $i_j$s are independent and zero mean. Further, assuming $w_j\in\{0,1\}$ are Bernoulli distributed with parameter $0.5$, we obtain:
\begin{align}
\sigma_\theta^2&=\mathbb{E}[ v^2_{\theta} ] = \frac{1}{C^2}\sum_{j=1}^N \mathbb{E}[w^2_j] \mathbb{E}[T_j] \frac{4}{3} g_m k T \\
&=\frac{1}{C^2}\sum_{j=1}^N \frac{1}{2} \frac{T_{\text{max}}}{2} \frac{4}{3} g_m k T = \left(\frac{T_{\text{max}} N }{C^2}\right)\frac{g_m k T}{3}
\end{align}
\color{black}
\noindent\textbf{Derivation of $\sigma_{\eta_\text{h}}^2$ in CM:}\\
We write the headroom clipping noise in CM as: %\cs{someone had made weird changes here - below is fixed. Thx}:
\begin{align}
    \eta_\text{h} &= %g(y_{\text{o}})-y_\text{o}%\nonumber\\
            \sum_{j=1}^N (w_j-\min (|w_j|,w_{\text{h}})\text{sign}(w_j))x_j %\nonumber\\
            =\boldsymbol{\lambda}^T\mathbf{x}
\end{align}
where $w_{\text{h}}=k_{\text{h}}\Delta_w$ is the smallest value of $|w_j|$ that leads to clipping, and $\boldsymbol{\lambda}$ is the clipping noise vector. The clipping noise terms can be assumed to be independent from each other and from the inputs. Furthermore, by virtue of the weights having a symmetric distribution, the clipping noise has zero mean, so that:
\begin{align}
\sigma_{\eta_\text{h}}^2 = N\mathbb{E} \left[ x^2 \right] \mathbb{E} \left[ \lambda^2 \right]. \label{eqn:clipvar_cm1}
\end{align}
In addition, the headroom clipping noise term variance is given by: $$\mathbb{E} \left[ \lambda^2 \right] = \text{Pr}\left(|w|\geq w_{\text{h}}\right) \mathbb{E} \left[ (w-w_\text{h})^2 \big| |w|>w_\text{h} \right]$$
Then, we use the bound $\text{Pr}\left(|w|\geq w_{\text{h}}\right)\leq\frac{\sigma^2_w}{w_{\text{h}}^2}$ by virtue of Chebyshev's inequality, and we evaluate $\mathbb{E} \left[ (w-w_\text{h})^2 \big| |w|>w_\text{h} \right] = \frac{(1-w_{\text{h}})_+)^2}{3}$.
Substituting into the above, we obtain an estimate for the headroom clipping noise term variance:
\begin{align}
\mathbb{E} \left[ \lambda^2 \right] \approx   \frac{1}{12} \sigma^2_wk_{\text{h}}^{-2}2^{2B_w}\left(1-2k_{\text{h}}2^{-B_w}\right)_+^2\label{eqn:apx-lamvar}
\end{align}
which we plug into \eqref{eqn:clipvar_cm1} to obtain the expression for the total clipping noise variance $\sigma_{\eta_\text{h}}^2$ in CM as listed Table \ref{tab:arch_summary}.
\noindent\textbf{Derivation of $\sigma_{\eta_\text{e}}^2$ in CM:}\\
We write the electrical noise in CM as:
$\eta_\text{e} = \sum_{j=1}^Nx_j\delta_{w_j}$ where $\delta_{w_j}$ is the electrical noise term corresponding to the discharge of weight $w_j$.
By virtue of inputs being independent from electrical noise terms, and assuming the latter are identically distributed, we have:
\begin{align}
\sigma_{\eta_\text{e}}^2 = N\mathbb{E} \left[ x^2 \right]\mathbb{E}[\delta^2_{w_j}].
\label{eqn:elec_var_cm1}
\end{align}
Next we derive $\mathbb{E}[\delta^2_{w_j}]$.
Though CM uses both QS and QR compute models, the noise source from QR dominates. Specifically the noise due to current mismatch (\ref{eqn:current-mistmatch}) dominates all other noise sources. In the presence of current mismatch noise, assuming the weight $w_j$ is positive, the discharge on the $j$-th BL in CM is given by:
\begin{align}
\Delta V_{\text{BL}_j}+\delta_{v_j}= \sum_{i=1}^{B_w-1}2^{-i} T_\text{max} \hat{w}_{i,j}(I_{i,j} + i_{i,j}) \label{eqn:apex-CM-discharge}
\end{align}
where $i_{i,j}$ is the noise due to current mismatch with noise variance given by (\ref{eqn:current-mistmatch}), $T_\text{max}=2^{B_w-1} T_\text{o}$, and $T_\text{o}$ is the smallest WL pulse. Therefore, the effective weight represented  the BL discharge is given by:
\begin{align}
&\frac{\Delta V_{\text{BL}_j}+\delta_{v_j}}{2^{B_w-1}\Delta V_{\text{BL},\text{unit}}}= \sum_{i=1}^{B_w-1} 2^{-i}\hat{w}_{i,j}\left(1 + \frac{i_{i,j}}{I_{i,j}}\right)\nonumber\\
&= w_j +\sum_{i=1}^{B_w-1} 2^{-i}\hat{w}_{i,j}\left( \frac{i_{i,j}}{I_{i,j}}\right)= w_j + \delta_{w_j} 
\label{eqn:apx-wnoise}
\end{align}
Note that the above equations assume $w_j$ is positive, similar equantions can be obtained if the weight is negative, where the discharge on BLB is considered instead of discharge on BL.
Thus we obtain:
\begin{align}
\mathbb{E}[\delta^2_{w_j}] = \sum_{i=1}^{B_w-1}4^{-i}\mathbb{E}[\hat{w}_{i,j}^2]\sigma_\text{D}^2 \label{eqn:apex-noise-var-w}
\end{align}
where $\sigma_\text{D}=\sigma_{I_j}/I_j$ ($\sigma_{I_j}$ is obtained using (\ref{eqn:current-mistmatch})). Assuming weight bits are equally likely to be 0 or 1, the above simplifies to:
$\mathbb{E}[\delta^2_{w_j}] = \frac{2\sigma_\text{D}^2}{3} \left(\frac{1}{4}-4^{-B_w}\right)$, which we plug into \eqref{eqn:elec_var_cm1} to obtain the expression for $\sigma^2_{\eta_{\text{e}}}$ in CM as listed in Table \ref{tab:arch_summary}.

\textbf{Derivation of $V_{\text{c}}$ in CM:}\\
In CM, BL discharge (\ref{eqn:apex-CM-discharge}) is:
$
\Delta V_{\text{BL}}=2^{B_w-1}V_{\text{BL},\text{unit}}w_i,
$ 
and therefore after multiplications and aggregations via charge sharing voltage at the input of the ADC is given by:
\begin{align}
\Delta V_{o}=\frac{2^{B_w-1}V_{\text{BL},\text{unit}}}{N}\sum_i^N w_ix_i
\end{align}
Therefore,
$
V_{\text{c}}=4\frac{2^{B_w}V_{\text{BL},\text{unit}}}{N}\sigma_y.
$ 
Since $\sigma_y=\sqrt{N}\mathbb{E}[x^2]\sigma^2_w$, we obtain the expression for $V_{\text{c}}$ in CM as listed in Table \ref{tab:arch_summary}.
%\begin{align}
%V_{\text{c}} = \frac{4\sigma_w 2^{B_w}\Delta V_{\text{BL},\text{unit}} \sqrt{\mathbb{E}\left[ x^2 \right]}}{\sqrt{N}}
%\end{align}

\textbf{Derivation of $\sigma_{\eta_\text{h}}^2$ in \QSArch{}:}\\
We write the headroom clipping noise in \QSArch as:
\begin{align}
    \eta_\text{h} &= %g(y_{\text{o}})-y_\text{o}%\nonumber\\
            \sum_{i=1}^{B_w}\sum_{j=1}^{B_x}2^{1-i-j} \lambda_{i,j}
\end{align}
where $\lambda_{i,j}$ is the headroom clipping noise term for every bit-wise DP. Note that the nature of two's complement arithmetic makes the overall headroom clipping noise zero-mean in spite of the individual headroom clipping noise terms being non zero-mean. Further, cross-correlations of headroom clipping noise terms are neglected, so that the total headroom clipping noise variance is given by:
$$\sigma_{\eta_\text{h}}^2 = \sum_{i=1}^{B_w}\sum_{j=1}^{B_x}4^{1-i-j}\mathbb{E}\left[\lambda_{i,j}^2\right]
$$
In addition, for independent, identically distributed headroom clipping noise terms, we obtain:
\begin{align}\sigma_{\eta_\text{h}}^2 = \frac{4}{9} \mathbb{E}\left[\lambda^2\right]\left(1-4^{-B_w}\right)\left(1-4^{-B_x}\right)
\label{eqn:clipvar_qsarch}
\end{align}
where the clipping noise term is 
$\lambda = (y_{\text{BL}}-y_{\text{clip}})\mathbbm{1}_{\{y_{\text{BL}}>y_{\text{clip}}\}}$ 
with $y_{\text{BL}}$ being the discharge on a bit-line per bit-wise DP. 
Assuming weight and input bits to be independent and equally likely to be 0 or 1, we obtain that $y_{\text{BL}}$ follows a binomial distribution $Bi(\frac{1}{4})$ so that:
\begin{align*}\mathbb{E}\left[\lambda^2\right] = \sum_{k=k_{\text{h}}}^N \left(k-k_{\text{h}}\right)^2\binom{N}{k} \left(\frac{1}{4}\right)^k\left(\frac{3}{4}\right)^{N-k}\end{align*}
%and thus we get the total clipping noise as follows:
%\begin{align}
%\sigma^2_{\eta_{\text{h}}} = \frac{4}{9}&\left(1-4^{-B_w}\right)\left(1-4^{-B_x}\right)\nonumber\\
%&\sum_{k=k_{\text{h}}}^N \left(k-k_{\text{h}}\right)^2\binom{N}{k} \left(\frac{1}{4}\right)^k\left(\frac{3}{4}\right)^{N-k}
%\end{align}
which we plug into \eqref{eqn:clipvar_qsarch} to obtain the expression for $\sigma_{\eta_\text{h}}^2$ in \QSArch{} as listed in Table \ref{tab:arch_summary}.

\textbf{Derivation of $\sigma_{\eta_\text{e}}^2$ in \QSArch{}:}\\
We write the electrical noise in \QSArch as:
$$\eta_\text{e} = \sum_{i=1}^{B_w}\sum_{j=1}^{B_x}\sum_{k=1}^{N}2^{1-i-j} \delta_{i,j,k}$$
where $\delta_{i,j,k}$ is the electrical noise term due to circuit non-idealities which occurs when accessing the bit-cell at location $(i,k)$ during the $j^{\text{th}}$ cycle. By virtue of independence of electrical noise terms, we obtain the total electrical noise variance as  
\begin{align*}\sigma_{\eta_\text{e}}^2 = \sum_{i=1}^{B_w}\sum_{j=1}^{B_x}\sum_{k=1}^N4^{1-i-j} \text{Var}(\delta_{i,j,k})
\end{align*}
Further, for identically distributed electrical noise terms, the above simplifies to:
\begin{align}\sigma_{\eta_\text{e}}^2 =\frac{4}{9} N\left(1-4^{-B_w}\right)\left(1-4^{-B_x}\right)\text{Var}(\delta)
\label{eqn:elec_var_qsarch}
\end{align}
where $\delta$ is the electrical noise per bit-cell discharge whose variance is:
$$\text{Var}(\delta) = \frac{1}{4}\sigma_\text{D}^2,$$ 
with the $\frac{1}{4}$ term being due to the necessity of both input and weight bits to equal 1. This value of $\text{Var}(\delta)$ is plugged into \eqref{eqn:elec_var_qsarch} to obtain the expression for $\sigma^2_{\eta_{\text{e}}}$ in \QSArch{} as listed in Table \ref{tab:arch_summary}.

\textbf{Derivation of $V_{\text{c}}$ in \QRArch{}:}\\
In \QRArch{}, we estimate binary-weighted DP in each column using charge sharing, as:
$
V_i=\frac{V_{\text{dd}}}{N}\sum_j^N x_{j}w_{i,j}.
$
Since in $V_i$ is the ADC input in \QRArch{} we need its standard deviations to estimate $V_{\text{c}}$. Since $w_{i,j}$ is binary-valued, 
$
\mathbb{E}[V]=V_{\text{dd}} E[x]=0.5V_{\text{dd}} \mu_x, 
$ and
$
\mathbb{E}[(V-0.5V_{\text{dd}}\mu_x)^2]=\frac{V^2_{\text{dd}}}{4N} (2E[x^2_{j}]-\mu_x^2).
$
Since $V_{\text{c}}=8\sqrt{\mathbb{E}[V^2]-\mathbb{E}[V]^2}$, we obtain the expression for $V_{\text{c}}$ in \QRArch{} as listed in Table \ref{tab:arch_summary}.
%\begin{align}
%    V_{\text{c}}=4V_{\text{dd}}\sqrt{\frac{\mathbb{E} \left[ x^2 \right]+\sigma_x^2}{N}}
%\end{align}

\bibliographystyle{IEEEtran}
\bibliography{references}

\end{document}